\documentclass[epsf,aps,draft,twocolumn,showpacs,preprintnumbers,amsmath,amssymb,footinbib]{revtex4}
\usepackage{epsf}
\newcommand{\beq}{\begin{equation}}
\newcommand{\eeq}{\end{equation}}
\newcommand{\tr}{{\rm tr}\;}
\newcommand{\boldl}{{\bf L}}
\newcommand{\boldlt}{{\bf L}_{ 3}}
\newcommand{\ellN}{\ell_{N}}
\newcommand{\calp}{{\cal P}}
\newcommand{\calc}{{\cal C}}

\newcommand{\vx}{\vec{x}}

\newcommand{\calz}{{\cal Z}}
\newcommand{\ellthree}{\ell_{ 3}}

\newcommand{\ellfund}{\ell_{N}}

\newcommand{\wideellfund}{\widetilde{\ell}_{{ N}}}
\newcommand{\wideellfundconj}{\widetilde{\ell}_{\overline{{ N}}}}

\newcommand{\boldlN}{{\bf L}_{{N}}}

\newcommand{\boldladj}{{\bf L}_{{N^{2} - 1}}}
\newcommand{\elladj}{\ell_{{ N^{2} - 1}}}
\newcommand{\boldltwopm}{\bold{L}_{( N^{ 2}  \pm  N)/ 2}}
\newcommand{\elltwopm}{\ell_{( N^{ 2}  \pm  N)/ 2}}

\newcommand{\calr}{{\cal R}}

\newcommand{\sdiv}{{div}}
\newcommand{\tildl}{\widetilde{\ell}}
\newcommand{\deltilell}{\delta \widetilde{\ell}}
\def\anp#1#2#3{Annals Phys. {\bf #1}, #2 (#3)}
\def\atmp#1#2#3{Adv. Theor. Math. Phys. {\bf #1}, #2 (#3)}

\def\ibid#1#2#3{{\it ibid.} {\bf #1}, #2 (#3)}
\def\ijm#1#2#3{Intl. Jour. Mod. Phys. {\bf #1}, #2 (#3)}
\def\jhep#1#2#3{Jour. High Energy Phys. {\bf #1}, #2 (#3)}
\def\jmp#1#2#3{Jour. Math. Phys. {\bf #1}, #2 (#3)}
\def\jpg#1#2#3{Jour. Phys. G {\bf #1}, #2 (#3)}

\def\npb#1#2#3{Nucl. Phys. B {\bf #1}, #2 (#3)}
\def\npsb#1#2#3{Nucl. Phys. Proc. Suppl. B {\bf #1}, #2 (#3)}
\def\plb#1#2#3{Phys. Lett. B {\bf #1}, #2 (#3)}
\def\prc#1#2#3{Phys. Rev. C {\bf #1}, #2 (#3)}
\def\prd#1#2#3{Phys. Rev. D {\bf #1}, #2 (#3)}
\def\prl#1#2#3{Phys. Rev. Lett. {\bf #1}, #2 (#3)}
\def\phr#1#2#3{Phys. Rep. {\bf #1}, #2 (#3)}

\def\rpp#1#2#3{Rep. Prog. Phys. {\bf #1}, #2 (#3)}
\def\rmp#1#2#3{Rev. Mod. Phys. {\bf #1}, #2 (#3)}
\def\sjnp#1#2#3{Sov. Jour. Nucl. Phys. {\bf #1}, #2 (#3)}
\def\zpc#1#2#3{Z. Phys. C {\bf #1}, #2 (#3)}
\begin{document}
\preprint{MIT-CTP 3439}
\title{Deconfining Phase Transition as a 
Matrix Model of Renormalized Polyakov Loops}
\author{Adrian Dumitru,$^{a}$ Yoshitaka Hatta,$^{b,c,d}$ 
Jonathan Lenaghan,$^e$ Kostas Orginos,$^{f,g}$
and Robert D. Pisarski$^{d,h}$}
\affiliation{
$^a$Institut f\"ur Theoretische Physik, J.~W.~Goethe Univ.,
Postfach 11 19 32, 60054 Frankfurt, Germany\\
$^b$Department of Physics, Kyoto University, Kyoto 606-8502, Japan\\
$^c$The Institute of Physical and Chemical Research (RIKEN)
Wako, Saitama 351-0198, Japan\\
$^d$High Energy Theory, Dept. of Physics, Brookhaven National Lab., Upton,
NY, 11973, U.S.A.\\
$^e$Dept. of Physics, Univ. of Virginia, Charlottesville, VA, 22904, U.S.A.\\
$^f$RIKEN/BNL, Dept. of Physics, Brookhaven National Lab., Upton, N.Y., 11973,
U.S.A.\\
$^g$Center for Theoretical Physics, 
Laboratory for Nuclear Science and Department of Physics,
Massachusetts Institute of Technology,
Cambridge, MA 02139-4307\\
$^h$Niels Bohr Institute, Blegdamsvej 17, 2100 Copenhagen, Denmark \\
}
\begin{abstract}
We discuss how to extract renormalized from bare Polyakov loops 
in $SU(N)$ lattice gauge theories at nonzero temperature.
Single loops in an irreducible representation
are multiplicatively renormalized, without mixing,
through mass renormalization.
The values of renormalized loops in the four lowest
representations of $SU(3)$ were measured numerically
on small, coarse lattices.  We find that in magnitude, condensates
for the sextet and octet loops are approximately the square of 
the triplet loop.
This agrees with a large $N$ expansion, where
factorization implies that the
expectation values of loops in adjoint and higher representations 
are powers of fundamental and anti-fundamental loops.
The corrections to the large $N$ relations at three colors are
greatest for the sextet loop, $\sim 1/N$, and are found to be $\leq 25\%$.
The values of the renormalized triplet loop can be described
by a matrix model, with an effective action 
dominated by the triplet loop: the deconfining
phase transition 
for $N=3$ is close to the Gross--Witten point at $N=\infty$.
\end{abstract}
\date{\today}
\maketitle

\section{Introduction}

In a $SU(N)$ gauge theory, 't Hooft showed that the exact order
parameter for deconfinement is a global $Z(N)$ spin
\cite{zn,polyakov,gava,mclerran,sv_yaffe,banks_ukawa,resum,shuryak,susy1,susy2,susy3,susy4,susy6,susy7,susy8,susy9,kinetic,pol_loop_a,pol_loop_b,pol_loop_c,cargese,sannino,pol_loop_misc}.
The global $Z(N)$ symmetry arises topologically \cite{susy4}
from the center of the gauge group, and is exact in a pure gauge theory.
Quarks carry $Z(N)$ charge, and so break the gluonic $Z(N)$ symmetry.
Nevertheless,
numerical results from the lattice, 
termed flavor independence \cite{flavor_ind},
suggest that the gluonic $Z(3)$ symmetry may be an approximate symmetry of  
$QCD$.

At a nonzero temperature $T$, a gluonic
$Z(N)$ spin is constructed by starting with 
a thermal Wilson line, 
which wraps all of the way around in imaginary time.
The trace of the thermal Wilson line is 
the Polyakov loop \cite{polyakov},
\beq
\ellfund = \frac{1}{N} \; \tr \boldlN \; ,
\label{def_ellfund}
\eeq
and is gauge invariant.  This is the trace of the propagator
for an infinitely massive, test quark; 
the subscripts denote that the test quark is in
the fundamental representation, of dimension $N$.

As a gauge theory is heated, deconfinement occurs above a temperature $T_d$.
The confined phase is $Z(N)$ symmetric, so the expectation
value of the fundamental loop, which has unit 
$Z(N)$ charge, vanishes below $T_d$.
The gluon spin condenses in the deconfined phase, above $T_d$:
\beq
\left\langle \ellfund \right\rangle 
= e^{i \phi} \; \left|\left\langle \ellfund\right\rangle\right| \neq 0
\;\;\; , \;\;\; e^{i \phi N} = 1 \;\;\; , \;\;\; T > T_d \; ,
\label{cond_deconf}
\eeq
and thereby breaks the global $Z(N)$ symmetry spontaneously 
\cite{zn,polyakov,gava,mclerran,sv_yaffe,banks_ukawa,resum,shuryak,susy1,susy2,susy3,susy4,susy6,susy7,susy8,susy9,kinetic,pol_loop_a,pol_loop_b,pol_loop_c,cargese,sannino,pol_loop_misc}.  

To compute near $T_d$, it is necessary to employ
numerical simulations on the lattice 
\cite{flavor_ind,string_ten,four_colors,teper,sommer,lat_sb,lattice_review}.  
The difficulty is that 
the expectation value of the Polyakov loop 
is a bare quantity, and so suffers ultraviolet divergences.
This is due to an additive mass shift 
which the test quark undergoes with a lattice regularization.
In four spacetime dimensions,
the mass divergence for a test quark
is linear in the ultraviolet cutoff,
proportional to
the inverse of the lattice spacing, $a$.  
This mass renormalization affects the expectation
value of the bare Polyakov loop as the exponential of
a divergent mass, $m^\sdiv_N$, times the
length of the path
\cite{gervais_neveu,polyakov_ren,ren_other_a,ren_other_b,ren_cusps,ren_cusp_brem,ren_loop_old}:
\beq
\left|\left \langle \ellfund \right \rangle \right| \sim 
\exp \left(- \frac{m^\sdiv_N}{T} \right) \;\;\; , \;\;\;
m^\sdiv_N \; \sim \; \frac{1}{a} \; .
\label{divergence}
\eeq
For a Polyakov loop, the length of the path is $1/T$.

Gervais and Neveu \cite{gervais_neveu}, 
Polyakov \cite{polyakov_ren}, and others 
\cite{ren_other_a,ren_other_b,ren_cusps,ren_cusp_brem,ren_loop_old}
established that Wilson lines are renormalizable operators.
We review these results in sec. II, applying them to $\ell_\calr$, 
a single Polyakov loop
in an arbitrary, irreducible representation, $\cal R$.
A renormalized loop, $\widetilde{\ell}_\calr$,
is formed by dividing the bare loop
by the appropriate renormalization constant, $\calz_\calr$:
\beq
\widetilde{\ell}_\calr = \frac{1}{\calz_\calr} \; \ell_\calr \; \;\; , \;\;\;
\calz_\calr = \exp \left( - \frac{m^\sdiv_\calr}{T} \right) \; .
\label{rendloop}
\eeq
This is a standard type of mass renormalization \cite{zinn};
for example, in perturbation theory
$a m^\sdiv_\calr$ is a power series in the coupling constant.  
The only unusual feature is that because the Wilson
line is a nonlocal operator, the renormalization constant
depends upon the length of the path; in general,
the renormalization constant for a path of length $\cal L$ is
$\calz_\calr = \exp(- m^\sdiv_\calr {\cal L})$.

The real problem is how to extract the divergent masses
non-perturbatively.
In this paper we suggest a way of doing this.
Consider a set of lattices, 
all at the same physical temperature, $T$, but with different
values of the lattice spacing, $a$.
Since the number of time steps, $N_t=1/(aT)$, changes between
these lattices, the divergent mass, $a m^\sdiv_\calr$,
follows by comparing the values of the bare Polyakov loops.
This assumes that, as in perturbation theory,
$a m^\sdiv_\calr$ is a function only of the
temperature, and not of the lattice spacing, as $a \rightarrow 0$.
Given the renormalization constant $\calz_\calr$, 
the renormalized loop $\widetilde{\ell}_\calr$
then follows from (\ref{rendloop}),
up to corrections at finite lattice spacing $\sim a T$.

In an asymptotically free theory,
at high temperature the vacuum is trivial in perturbation theory, 
as the thermal Wilson line is a $Z(N)$ phase times the unit matrix,
$\boldlN \rightarrow e^{i \phi} \bold{1}_{ N}$ (\ref{cond_deconf}).  
After suitable normalization, 
the expectation value of any renormalized loop
approaches one at high temperature,
\beq
\left|
\left \langle \widetilde{\ell}_\calr \right \rangle \right|
\rightarrow 1
\;\;\; , \;\;\; T \rightarrow \infty \; .
\label{asymptotic}
\eeq

In sec. III we present measurements of bare and renormalized loops 
obtained through 
numerical simulations in a pure $SU(3)$ lattice gauge theory.
Polyakov loops in the four lowest representations were measured,
although we only found significant signals for three: the
fundamental, the symmetric two-index tensor,
and the adjoint representations.
For three colors, these are the triplet,
the sextet, and the octet representations, respectively.
These loops were measured
at temperatures from $\approx .5 T_d \rightarrow 3 T_d$.
Numerically, we find that in all representations, 
the divergent masses $a m^\sdiv_\calr$ are positive, so the bare
loops vanish in the continuum limit, $N_t \rightarrow \infty$.
The values of
renormalized loops appear to have a well-defined continuum limit
and approximately satisfy the asymptotic condition of (\ref{asymptotic}).

An alternate procedure for computing the renormalized Polyakov loop
was developed by
Kaczmarek, Karsch, Petreczky, and Zantow \cite{bielefeld_ren}.
They obtain $\calz_\calr$ from the 
two point function of Polyakov loops at short distances.
Their numerical values
for the triplet Polyakov loop agree with ours at $T_d$, but differ 
at higher temperatures; they did not consider higher representations.

The most basic thing to consider is the size of the
renormalized triplet loop.  As we shall see, 
loops in higher
representations are approximately given as powers of the
triplet loop.
We define a perturbative regime when
the expectation value of the renormalized triplet loop is near one
in magnitude.
If the triplet loop is nonzero, but not close to one, then we
have a deconfined, but non-perturbative, regime.
At the transition, $T=T_d^+$, both we and Kaczmarek {\it et al.} 
\cite{bielefeld_ren} find that the renormalized triplet loop,
$|\langle \widetilde{\ell}_3 \rangle|$, is $\approx .4$; by 
$\approx 3 T_d$, we find it is $\approx .9$, while \cite{bielefeld_ren}
finds $\approx 1.0$.  This suggests that
a pure gauge theory is, in some sense, perturbative from temperatures
of $\approx 3 T_d$ on up, but not from $T_d$ to $\approx 3 T_d$.
This is in qualitative agreement
with different resummations of perturbation theory \cite{resum}, all of
which work down only to a temperature which is several times $T_d$.
It is also suggested by some features of the RHIC data \cite{shuryak}.

Turning to loops in higher
representations, we find that
the expectation values of {\it all} renormalized loops are very
small in the confined phase.
Renormalized loops are nonzero above $T_d$, with an ordering
of expectation values as
triplet, octet, and then sextet loop.  Even though the octet and sextet
condensates are smaller than that for the triplet, they are 
still significant.

The apparently large values for the condensates of the
octet and sextet loops are illusory.
This is because when the fundamental loop condenses, that
alone induces expectation values for all higher loops.  It
is to these induced values that we must compare.

This is clear in the limit of an infinite number of colors \cite{susy2}.
Migdal and Makeenko observed that in $SU(N)$ gauge theories,
expectation values factorize at large $N$
\cite{largeN,eguchi_kawai,gocksch_neri,gross_witten,largeN_trick,mixed_action,mixed_action_2,kogut,matrix_deconfining,green_karsch,leutwyler_smilga,matrix_review,gross_taylor}.
Factorization is the statement that disconnected diagrams, with the
most traces, dominate at large $N$.  

At infinite $N$, factorization fixes
the expectation value of any Polyakov loop to be equal to powers
of those for the fundamental and anti-fundamental,
$\wideellfundconj=(\wideellfund)^*$, loop \cite{eguchi_kawai}:
\begin{equation}
\left\langle \widetilde{\ell}_\calr\right\rangle
= \left\langle \wideellfund \right\rangle^{p_+}
\left\langle \wideellfundconj \right\rangle^{p_-}
+ O\left(\frac{1}{N}\right) \;\;\; . \;\;\;
\label{mean_fieldA}
\end{equation}
Hence
\begin{equation}
\left\langle \widetilde{\ell}_\calr\right\rangle
= e^{i e_\calr\phi} \;
\left|\left\langle \wideellfund \right\rangle\right|^{p}
+ O\left(\frac{1}{N}\right) \; .
\label{mean_fieldB}
\end{equation}
The integers $p_+$ and $p_-$ are determined from the
Young tableaux of the representation $\calr$, using the composite
representations of Gross and Taylor \cite{gross_taylor,group}.
At any $N$, the overall phase is
fixed, trivially, by the $Z(N)$ charge of $\calr$,
$e_\calr \equiv p_+ - p_-$, modulo $N$.
What is not trivial is the magnitude of the loop: at large $N$, 
the term with the most powers of the fundamental loop
dominates, with power $p \equiv p_+ + p_-$.

Lattice simulations
with two colors by Damgaard and others \cite{damgaard,twocolors}
showed that the bare adjoint loop is an approximate order
parameter for deconfinement.  
At infinite $N$, by factorization any renormalized loop serves
as an order parameter for deconfinement, independent of
its $Z(N)$ charge.
For example, consider the adjoint loop, with
$p_+ = p_- = 1$, and the loop for the
symmetric two-index tensor representation,
$p_+ = 2$ and $p_- = 0$.  While the adjoint
has no $Z(N)$ charge, and the two-index tensor charge
two, modulo $N$, in magnitude both expectation values are
$\sim |\langle \widetilde{\ell}_N\rangle|^2$ at large $N$.

We tested these large $N$ relationships numerically for three
colors.  For each loop, we define the difference between
the measured loop and its value in the large $N$ limit.  The 
expectation value of the sextet difference is defined to be
\beq
\left \langle \delta \widetilde{\ell}_{ 6}\right \rangle
= \left \langle \widetilde{\ell}_{ 6}\right \rangle -
\left \langle\widetilde{\ell}_{ 3}\right\rangle^{ 2} \; ,
\label{sextet_diff}
\eeq
and that for the octet difference as
\beq
\left\langle \delta \widetilde{\ell}_{ 8} \right\rangle
= \left \langle \widetilde{\ell}_{ 8}\right\rangle -
\left|\left\langle \widetilde{\ell}_{ 3}\right\rangle\right|^2 \; .
\label{octet_diff}
\eeq
Of course there is some ambiguity in defining the difference loops.
One advantage of the above definitions is that they automatically
vanish at both low and high temperature.
In the confined phase, $T \leq T_d$, the difference loops (nearly) vanish
because all loops are (essentially) zero; at very high
temperature, the difference loops vanish because all loops
approach one as $T \rightarrow \infty$.

The expectation values of the difference loops
show interesting behavior.  
They vanish below $T_d$, and spike down above $T_d$,
with a maximum at a temperature $> T_d$.
The spike for the octet difference is smaller,
narrower, and closer to $T_d$ than the spike for the sextet
difference:
$|\langle\delta \widetilde{\ell}_{{\bf 8}}\rangle| \leq .2$, with
a maximum at $\approx 1.1 T_d$, while
$|\langle\delta \widetilde{\ell}_{{\bf 6}}\rangle| \leq .25$, 
with a maximum at $\approx 1.3 T_d$.

The magnitude of these expectation values
are in accord with a large $N$ expansion.
Corrections to the sextet difference are larger,
$\delta \widetilde{\ell}_{{ 6}} \sim 1/N$,
than those for the octet difference,
$\delta \widetilde{\ell}_{{ 8}} \sim 1/N^2$.
Thus our measurements of the values of renormalized
loops give us a numerical estimate of just how
good a large $N$ approximation is for three colors.
Corrections to the sextet difference, of $\sim 1/N$, are
found to be $\leq 25\%$ when $N=3$.  

Although factorization tells us how to reduce condensates
for higher loops to powers of that for the fundamental loop,
it does not tell us how the condensate for the fundamental loop
changes with temperature.  
Given the mean field relations satisfied
by loops in higher representations, in sec. IV we consider
a mean field theory for the fundamental loop itself.  
We consider a matrix valued mean field theory, or matrix model;
this arises in a wide variety of contexts
\cite{eguchi_kawai,gocksch_neri,gross_witten,largeN_trick,mixed_action,mixed_action_2,kogut,matrix_deconfining,green_karsch,leutwyler_smilga,matrix_review,gross_taylor},
including previous 
\cite{matrix_deconfining,green_karsch,damgaard}
and recent
\cite{susy7,susy8,susy9}
work on the deconfining transition.

The most general effective action for a matrix model of the deconfining
transition involves a sum over loops in all
representations.  For three colors, 
we find that the lattice data for the renormalized triplet loop
is approximately described by a model whose action
includes only the triplet loop.
Over the range of temperatures studied, the coupling constant
for the triplet loop is nearly linear in temperature. 
While the overall values
of the sextet and octet loops are approximately described by
this mean field theory, the difference loops are not.
We categorize more involved models which might.

It is interesting to consider what matrix models
might apply to the deconfining transition for more than three colors.
Consider the simplest possible model, where the action includes
just the fundamental loop.  At infinite $N$, the solution
follows from that of Gross and Witten 
\cite{susy7,susy8,susy9,gross_witten,kogut,green_karsch,matrix_review,damgaard}.
Kogut, Snow, and Stone showed that in this model,
the deconfining transition is of first order, with a latent heat $\sim N^2$ 
\cite{susy7,susy8,kogut,green_karsch,matrix_review,damgaard}.  
As a first order transition, the fundamental loop 
jumps, from zero to precisely one half:
\beq
\left|\left\langle \widetilde{\ell}_{{ N}}
\right\rangle\right| = \frac{1}{2} 
\;\;\; , \;\;\; 
T = T_d^+ 
\;\;\; , \;\;\; 
N=\infty \; .
\label{vev_Td}
\eeq
This was also stressed recently by 
Aharony, Marsano, Minwalla, Papadodimas, and Van Raamsdonk \cite{susy7}:
at $T_d$, this theory only deconfines {\it halfway}.

It is striking that this special value at large $N$ 
is close to that found from numerical
simulations for the renormalized, triplet loop; both we
and Kaczmarek {\it et al.} \cite{bielefeld_ren} find 
$|\langle \widetilde{\ell}_3 \rangle | \approx .4$ at $T=T_d^+$.
In the $N=3$ matrix model closest to the Gross--Witten point,
this value is not $\frac{1}{2}$, but $\approx .485 \pm .001$ 
\cite{kogut,damgaard,coming1}.
The value in the Polyakov loop model, which we discuss shortly,
is $\approx .55$ \cite{pol_loop_b}.

The first order transition at the Gross--Witten point
is atypical.
In ordinary first order transitions, masses are nonzero on either
side of the transition  \cite{zinn}.
Even though the value of the 
fundamental loop jumps at $T_d$, at the Gross--Witten point
both the string tension, and a gauge invariant
Debye mass, vanish.
This is only possible because of a transition, which is of third order
in the matrix model coupling constant, at infinite $N$ \cite{gross_witten}.
The Gross--Witten point is specific to infinite $N$:
at finite $N$, but $\geq 3$, in the matrix model deconfinement is an
ordinary first order transition, with 
the string tension and the Debye mass nonzero at $T_d$.

For three colors, lattice simulations find a 
relatively weak first order transition,
accompanied by a large decrease in both
the string tension and the Debye mass near $T_d$,
each by about a factor of ten \cite{string_ten}.
The customary explanation 
for this is that, as in the Potts model,
three colors is near the
second order transition known to occur for two colors \cite{twocolors}.  
We suggest that the deconfining phase transition for three colors
is also close to the Gross--Witten point of infinite $N$;
exactly how close can be categorized in a matrix model \cite{coming1}.

Aharony {\it et al.} \cite{susy7}
showed how at large $N$ the Hagedorn temperature can be
computed when space is a very small sphere.
They find that the 
deconfining transition is of first order if the Hagedorn temperature
is greater than $T_d$.  It is tempting to think that the spikes
which we found for 
the sextet and octet difference loops may be related to the Hagedorn
temperature.  If so, for three colors the Hagedorn
temperature is tens of percent above that for the deconfining transition.

It would be valuable to know from numerical simulations
if the deconfining transition for more than three colors is
close to the Gross--Witten point as well, or if that 
is unique to three colors.

An Appendix gives a formal discussion of
improved Wilson lines; on the lattice, these are related to
smeared, stout links \cite{improved}.  This may be of use for
measuring Polyakov loops in higher representations.

Our work was motivated by the Polyakov loop model, which postulates
a relationship between the Polyakov loop --- which was
presumed to exist as a renormalized quantity --- and the
pressure \cite{pol_loop_a,pol_loop_b,pol_loop_c,cargese}.  
In the end, we have more than expected:
not just a renormalized Polyakov loop, but
a good approximation to its potential, in the $SU(3)$ matrix model.

\section{Bare and Renormalized Polyakov Loops}

\subsection{Traces of Wilson lines in imaginary time}

At a temperature $T$, the thermal Wilson line at a spatial
point $\vec{x}$, running in time from 
$0$ to $\tau$, is
\beq
\boldl_\calr(\vec{x},\tau) = \calp \; \exp 
\left(  i g \int^{\tau}_0 A_0^a(\vec{x},\tau') \;
{\bf t}^a_\calr \; d \tau' \right)  \; ;
\eeq
we take the representation $\calr$ to be irreducible.
The notation is standard: $\calp$ denotes path ordering, 
$g$ is the gauge coupling constant, 
$\tau$ and $\tau'$ are variables for
imaginary time, $\tau, \tau': 0 \rightarrow 1/T$,
$A_0^a$ is the vector potential in the time direction, and
${\bf t}^a_\calr$ are the generators of $SU(N)$ in $\calr$.

The Wilson line is a $SU(N)$ phase factor for
$\calr$, and so is a unitary matrix,
\beq
\boldl_\calr^\dagger(\vx,\tau) \;
\boldl_\calr(\vx,\tau) = {\bold 1}_{d_\calr} \; ;
\label{unitary_relation}
\eeq
$d_\calr$ is the dimension of the representation, and
${\bold 1}_{d_\calr}$ the unit matrix in that space.

The thermal Wilson line is proportional to the propagator of a ``test'' quark
in the representation $\calr$
\cite{ren_other_a,cargese}.  A test quark is one whose mass
is so large that if you put it at a given point in space, $\vx$, it
just sits there.  The only motion of a test quark is up
in imaginary time.  While the test quark doesn't move in space,
it still interacts in color space: through the
Aharanov-Bohm effect, it acquires a $SU(N)$ phase.

To see this, form a covariant
derivative in imaginary time, and define a propagator, ${\cal G}_\calr$, as
its inverse:
\beq
\left( \frac{d}{d\tau} \; {\bf 1}_{d_\calr} - i \, g \, A^a_0(\vx,\tau) 
\, {\bf t}^a_\calr \right)
\; {\cal G}_\calr(\vx,\tau) \; = \; \delta(\tau) \; {\bf 1}_{d_\calr}  \; .
\label{propagator}
\eeq
It is easy computing the propagator in one dimension: it
is just a step function, $\theta(\tau)$, times the Wilson line, 
\beq
{\cal G}_\calr(\vx,\tau) = \theta(\tau) \; \boldl_\calr(\vx,\tau) \; .
\eeq

Alternately, consider the path integral representation for
the propagator of a particle with mass $m$ in a background gauge
field; schematically,
\beq
\int {\cal D}x^\mu \; \exp \left( - \int \left(
m \sqrt{\dot{x}^2} + i g A_\mu \dot{x}^\mu \right) ds \right)\; ,
\label{path_integral}
\eeq
where $\dot{x}^\mu = dx^\mu/ds$, and $s$ is the path length;
an exact form is given in \cite{gervais_neveu}.
In the limit of $m\rightarrow \infty$, this path is a straight
line up in imaginary time, and this propagator is the thermal Wilson line.
Classically, the partition function
in (\ref{path_integral}) is $\sim \exp(- m {\cal L})$, where
$m$ is the bare mass, and $\cal L$ is the length of the path.

Under a gauge transformation in $\cal R$,
$\Omega_\calr(\vx,\tau)$, the Wilson line
transforms as 
\beq
\boldl_\calr(\vx,\tau) 
\rightarrow \Omega_\calr^\dagger(\vx,\tau) \boldl_\calr(\vx,\tau)
\Omega_\calr(\vx,0) \; .
\eeq
As bosons, the gauge fields are
periodic in imaginary time, with period $1/T$.
For the time being, we also assume that all gauge
transformations are periodic in time,
$\Omega_\calr(\vx,1/T)=\Omega_\calr(\vx,0)$.  We 
relax that assumption later, but only in a way which affects
the global symmetry, and not the local symmetry.

For periodic gauge transformations, we can form a quantity which
is locally gauge invariant by wrapping the Wilson line around in imaginary
time, and then taking its trace:
\beq
\tr \boldl_\calr(\vx,1/T) \; .
\label{wrap_all}
\eeq
For the time being we follow the custom of mathematics, which is 
to work with traces which are not normalized.  
The trace is greatest when the Wilson line is the identity,
$= d_\calr$.

The propagation of a test quark, at $\vx$, forward in imaginary
time generates the Wilson line in the fundamental representation,
$\boldl_{{ N}}(\vx,1/T)$.  A test anti-quark is a test quark
moving backward in imaginary time, so it gives
the conjugate Wilson line,
$\boldl_{\overline{{ N}}}(\vx,1/T) 
= \boldl_{{ N}}^\dagger(\vx,1/T)$.

Let us consider how to combine more test quarks and anti-quarks.
To be gauge invariant, 
the Wilson lines must wrap around completely in imaginary time, so
we drop the dependence on imaginary time, $1/T$.
We also assume that all test quarks and anti-quarks
are put down at the same point in space, and so drop the dependence
on $\vx$ as well. 
So a test quark gives $\boldlN$, and
a test anti-quark, $\boldlN^\dagger$.

The general classification of representations is, for arbitrary
$N$, rather involved \cite{gross_taylor,group}.  We thus
start with the lowest representations for general $N$.
We then discuss some simplifications for $N=3$.
Finally, we show how one can easily classify all representations
in the large $N$ limit \cite{gross_taylor}.  We use
a notation where $\calr$ is generally denoted by its
dimension, $d_\calr$.

\subsubsection{Simple examples}

The adjoint Wilson line is a test meson, constructed
from a test quark and anti-quark.  To combine 
the fundamental and anti-fundamental Wilson lines into
something with adjoint indices, we sandwich
$\boldlN$ and $\boldlN^\dagger$
between two $SU(N)$ generators, $t^a_{{ N}}$,
\beq
\boldladj^{ab} \; = \; \tr \left( \boldlN \; t^a_{{ N}} \;
\boldlN^\dagger \; t^b_{{ N}} 
\right) \; ;
\label{adja}
\eeq
the adjoint indices $a,b=1\ldots(N^2-1)$.  
The trace of the adjoint Wilson line is
\beq
\tr \boldladj =
| \tr \boldlN |^2 - 1 \; .
\label{adjb}
\eeq
This follows from an identity on the $t^a$'s, or directly from
group theory.  
The product of the fundamental
and anti-fundamental representations is the sum of
the adjoint and identity, so the
coefficient of $|\tr \boldlN|^2$ in (\ref{adjb}) is
one.  To check the $-1$ in (\ref{adjb}), consider the
case when the Wilson line is the unit matrix, 
$\boldlN = {\bold 1}_{{ N}}$; then the adjoint
trace is its dimension, $= N^2 - 1$.

The next representations are tensors with two fundamental indices
\cite{group}.  These represent the propagation of two test quarks
up in imaginary time.  Two Wilson lines can be put together in
either a symmetric
($+$), or an anti-symmetric ($-$), way.  The Wilson lines for
the two-index representations are
\beq
\boldltwopm^{ij;kl}
= \frac{1}{2} \left( \boldl^{ik}_{{ N}} 
\boldl^{jl}_{{ N}} 
\pm \boldl^{jk}_{{ N}} \boldl^{il}_{{ N}}
\right) \; .
\eeq
Here $i,j,k,l = 1 \ldots N$ are indices for the fundamental representation.
The traces of the representations with two Wilson lines are then
\beq
\tr \boldltwopm
= \frac{1}{2} \left( \left( \tr \boldlN \right)^2 
\pm \tr \boldlN^2 \right) \; .
\label{sym_anti}
\eeq
Checking the coefficients for $\boldlN = {\bold 1}_{{ N}}$,
the dimensions of the representations are as indicated.
The first term is a product of two Wilson lines, each of which wrap
around once in imaginary time.  The second term is one Wilson
line which wraps around twice in imaginary time.  

\subsubsection{Three colors}

For three colors, the fundamental representation is a triplet,
${\bold 3}$,
the adjoint representation is an octet, $\bold 8$,
and the symmetric two-index representation is a sextet, $\bold 6$.

Special to three colors, the anti-symmetric two-index
representation is the anti-triplet,
$\overline{{\bold 3}}$.  To see this,
diagonalize the triplet Wilson line by a 
local gauge rotation.  After diagonalization,
each element of $\boldl_{{ 3}}$ is a phase; the
product of all phases is one:
\beq
\boldl_{3} = 
\left(
\begin{array}{ccc}
\exp(i \alpha_1) & 0                & 0                             \\
0                & \exp(i \alpha_2) & 0                             \\
0                & 0                & \exp(- i(\alpha_1 + \alpha_2))\\
\end{array}
\right) \; .
\label{diag}
\eeq
Then it is easy to check that 
$\boldl_{({ N^2-  N)/ 2}}$
from (\ref{sym_anti}) $=\tr \boldl^\dagger_{ 3}$
when $N=3$.  Notice that (\ref{sym_anti}) actually gives
the anti-triplet loop, which is a fault of our notation.

We mention one other representation for three colors.
Consider a test baryon, composed of three test quarks.
The symmetric combination of three fundamental Wilson lines
gives the decuplet representation when $N=3$.  Its trace is
\beq
\tr \boldl_{{ 10}} = 
\frac{1}{6} \left(
\left( \tr \boldlt \right)^3 
+ 3 \; \tr \boldlt \; \tr \boldlt^2 + 2 \; \tr \boldlt^3 
\right) \; .
\label{decup}
\eeq
Since $\boldlt$ is a $SU(3)$ matrix, 
$$
\det \boldlt = 1
$$
\beq
= \frac{1}{6} \; \left(\tr \boldlt \right)^3 - \frac{1}{2} \; \tr \boldlt \; \tr \boldlt^2 + \frac{1}{3} \; \tr \boldlt^3
\; ,
\label{det}
\eeq
which is Mandelstam's constraint 
\cite{mandelstam}.  Using it, we find that the trace of the
decuplet Wilson line is
\beq
\tr \boldl_{{ 10}}
= \tr \boldlt \; \tr \boldlt^2 + 1 \; .
\label{decup_two}
\eeq

\subsubsection{Large $N$}

The usual classification of representations is given using
Young tableaux \cite{gross_taylor,group}.  This is not
very convenient for the large $N$ limit, though.  The
reason is elementary: Young tableaux involve the 
construction of tensors with fundamental indices.
While of course complete, it does not naturally incorporate
the symmetry between the fundamental and anti-fundamental
representations.  This is accomplished using the composite
representations of Gross and Taylor \cite{gross_taylor};
we give an abbreviated summary which is sufficient for our purposes.

Consider forming a state with $p_+$ test quarks and $p_-$ test
anti-quarks.  This is done by combining $p_+$ fundamental
Wilson lines, and $p_-$ anti-fundamental Wilson lines, in all
possible ways.  To be gauge invariant, we must take
traces of $\boldlN$ and $\boldlN^\dagger$.  We also
have to remember that 
$\boldlN$ is a unitary matrix,
$\boldlN^\dagger \boldlN = {\bold 1_{ N}}$, (\ref{unitary_relation}).
This relation implies that all traces are either traces
of powers of $\boldlN$, or traces of powers of $\boldlN^\dagger$,
separately.  Any mixed terms can be simplified using the unitary
relation.

The explicit construction of Wilson lines
in different representations, as done above
for the adjoint and the two-index tensor, is unnecessary.
Only gauge invariant quantities matter: these are
traces of the Wilson line in different representations.
For any $\calr$, by the Frobenius formula \cite{gross_taylor,group}
we can express the trace of a Wilson line in $\calr$
as sums of products of traces of the fundamental Wilson line.

We do this naively, by considering how 
to combine $p_+$ $\boldlN$'s 
and $p_-$ $\boldlN^\dagger$'s.  The simplest thing is to
take a trace of every Wilson line, separately:
\beq
(\tr \boldlN)^{p_+} \;
(\tr \boldlN^\dagger)^{p_-} \; .
\label{traceA}
\eeq
In this term, each Wilson
line wraps once, and only once, around in imaginary time.

This is just the first term in a long series, though.  
To start with, we can consider a term where one Wilson line
wraps around twice in imaginary time:
\beq
\; (\tr \boldlN)^{p_+-2} \; (\tr \boldlN^2) \;
(\tr \boldlN^\dagger)^{p_-} \; .
\label{traceB}
\eeq
Continuing, a term where one Wilson line wraps around thrice:
\beq
\; (\tr \boldlN)^{p_+ - 3} \; (\tr \boldlN^3) \;
(\tr \boldlN^\dagger)^{p_-} \; ,
\label{traceBb}
\eeq
and so on.
We can do this as well for the anti-Wilson lines; a Wilson
line which wraps around backward twice gives
\beq
(\tr \boldlN)^{p_+} \;
(\tr \boldlN^\dagger)^{p_- - 2} \; \tr (\boldlN^\dagger)^2 \; .
\label{traceC}
\eeq
We continue in this fashion.  The series continues down,
generating fewer traces overall, as they are replaced
by traces with higher powers of the Wilson and anti-Wilson line.
In general, the operator
$\tr \boldl^{p_+}$ represents a Wilson line which wraps
around $p_+$ times forward in imaginary time; $\tr (\boldl^\dagger)^{p_-}$,
a Wilson line which wraps around $p_-$ times backward.

The series stops when
we only have either one or no traces left.  
If $p_+ = p_-$, the series stops at a constant, with no trace.
If $p_+ \neq p_-$, the series stops at one trace,
\beq
\tr \; \boldlN^{p_+} \; (\boldlN^\dagger)^{p_-} \; .
\label{traceD}
\eeq
This term can obviously be reduced, as it is a Wilson line
which goes forward $p_+$ times, and then backward $p_-$ times.

The above series of traces of Wilson lines represents the propagation of
test quarks and anti-quarks.  All we can do with test quarks is
to take loops in imaginary time, so the only question left is
how many times a test quark, or anti-quark, goes around in
imaginary time.  

When $p_- = 0$ these terms are the
Schur functions \cite{gross_taylor,group}.  When
$p_- \neq 0$, these terms are the Schur functions for
the composite representations of \cite{gross_taylor}.

Consider now the powers of $N$ at large $N$.  We assume
that in the deconfined phase, each trace gives a power
of $N$: $\tr \boldlN^q \sim N$ for all $q$.
The first term in (\ref{traceA}) is of order 
$\sim N^p$.  Terms in (\ref{traceB}) and (\ref{traceC})
have one fewer trace overall, and so 
are $\sim N^{p-1}$, down by one power of $1/N$ relative
to (\ref{traceA}).  Each time a Wilson line wraps around
an extra time in imaginary time, a possible trace is lost,
which suppresses the term by $\sim 1/N$ relative to (\ref{traceA}).

This assumes that the coefficients of all terms,
(\ref{traceA}-\ref{traceD}), are of order one.
For example, if the coefficient of (\ref{traceB}) or (\ref{traceC})
were $\sim N$ relative to that of (\ref{traceA}), then we couldn't
conclude anything about the large $N$ limit.  Group theory
tells us, however, that this is not so: for any
$N$, all coefficients are numbers of order one \cite{gross_taylor,group}.
This is because all representations are constructed by 
symmetrization
and anti-symmetrization operators \cite{group};
the action of these operators 
depends on the number of indices, but not upon $N$.

Consequently, at large $N$ the trace of any representation is dominated
by the term where every Wilson line (or anti-line) wraps
around only once in imaginary time.  This is just because 
we maximize the number of possible traces, and so the
powers of $N$.
At large $N$ we denote representations as
$\calr =  N \wedge p$:
\beq
\tr \boldl_{N \wedge p}
\sim 
(\tr \boldlN)^{p_+} \;
(\tr \boldlN^\dagger)^{p_-} \; ,
\label{Nwedgep}
\eeq
$p=p_+ + p_-$.  The notation is meant to be
suggestive, as the dimension of this representation is 
$\sim N^{p}$ at large $N$.  

The integers $p_+$ and $p_-$ can be computed from the
Young tableaux of the representation \cite{gross_taylor}.  
Denote the columns of the Young tableaux by the index $i$,
and separate them into two categories.
If the number of rows in a column, $r^i$, is $\leq N/2$, then
we leave the column alone, and refer to it as a fundamental
column, with $r_+^i = r^i$ rows.  If the number of rows in a column is $> N/2$,
then we turn it into a column with $r^i_- = N-r^i$ anti-fundamental
rows.  Then $p_+ = \Sigma_i r_i^+$ 
is the number of boxes in all of the fundamental
columns, and $p_- = \Sigma_i r_i^-$ 
is the number of boxes in all of the anti-fundamental
columns.  In the limit of infinite $N$, no other details of
the Young tableaux matter; all that matters is the
total number of boxes $p_+$ and $p_-$.

In group theory, the distinction between fundamental and
anti-fundamental columns appears awkward.  It is essential
to understand the large $N$ limit, though.  
Consider, for example, a single anti-fundamental box in a
Young tableaux: this is given by a column with $N-1$ rows.
If we counted rows naively, 
this would give us $N-1$ powers of a trace.  Clearly we
should replace this by one anti-fundamental line, and one trace.
Geometrically, if a Wilson line wraps forward around $r_i$ times
in imaginary time, when
$r_i > N/2$ it is better to replace it by a Wilson line going
backward $N - r_i$ times.

While we do not need it now, we note that 
the Casimirs of the representations $\bf N \wedge \bf p$
were computed by Samuel \cite{mixed_action_2} and Gross
and Taylor \cite{gross_taylor}.
While the dimension grows strongly, like $N^p$,
the Casimir does not; at large $N$, it is linear
in $N$, proportional to $p_+ + p_-$:
\beq
{\cal C}_{\bf N \wedge \bf p} =
(p_+ + p_-) \; \frac{N}{2} + O(1) \; .
\label{casimir}
\eeq
We shall see in sec (II.D) that this
relation for the Casimir ensures that 
Polyakov loops, as computed to lowest order
in perturbation theory, satisfy factorization
at large $N$, (\ref{large_N_fact}) and (\ref{pert_loop}).

\subsection{$Z(N)$ charges and confinement}

Previously, we required that gauge transformations be
strictly periodic in imaginary time.  't Hooft noted that
this is not necessary.  Consider gauge transformations which
are periodic, in $1/T$, only up to a constant.
So as not to change the periodicity of the gauge fields,
this constant gauge transformation must be 
equal to the identity matrix times a phase,
equal to an $N^{th}$ root of unity, (\ref{cond_deconf}).
For the fundamental representation, aperiodic gauge transformations are
\beq
\Omega_{{ N}}(\vx,1/T) = 
e^{i\phi} \; \Omega_{{ N}}(\vx,0) 
\;\;\; , \;\;\;
e^{i \phi N} =1 \; .
\label{phase_1}
\eeq
This phase represents the center of the local $SU(N)$ gauge group,
which is a global group of $Z(N)$:
$\phi = 2 \pi j/N$, $j=0,1\ldots(N-1)$.

As the $Z(N)$ phases commute with any element of the group, 
the gauge fields remain strictly periodic
under such an aperiodic gauge transformation.
The Wilson line, however, does change:
for the fundamental representation,
\beq
\boldl_{{ N}} \rightarrow e^{i\phi} \; \boldl_{{ N}} \; .
\eeq

We define the $Z(N)$ charge, or the $N$-ality, of the fundamental
Wilson line to be one, $e_{{ N}} = 1$.
The traces of Wilson lines in other representations transform
under aperiodic gauge transformations as
\beq
\boldl_\calr \rightarrow e^{i e_\calr \phi} \; \boldl_\calr \; ;
\label{zNchg}
\eeq
the $Z(N)$ charge $e_\calr$ is an integer.

Due to the cyclic nature of $Z(N)$, charge is only defined modulo $N$.
If the fundamental
has charge one, the anti-fundamental has charge minus one,
which is equivalent to charge $N-1$.
The simplest field with vanishing $Z(N)$ charge is the adjoint.
A baryon Wilson line, such as the symmetric combination of $N$
fundamental indices, is also $Z(N)$ neutral.
At large $N$, the Wilson line in
the ${\bf N \wedge \bf p}$ representation of (\ref{Nwedgep})
has $Z(N)$ charge $e_\calr = p_+ - p_-$, modulo $N$.

For three colors, the anti-triplet and sextet
representations have charge two, which is the
same as minus one.  As a test baryon, the decuplet Wilson line
is $Z(3)$ neutral.

The confining phase, for $T \leq T_d$, 
is characterized by an unbroken global $Z(N)$ symmetry \cite{zn}.
Hence the traces of Wilson lines with nonzero $Z(N)$ charge 
vanish below $T_d$,
\beq
\left \langle \tr \boldl_\calr \right\rangle = 0 \;\; , \;\;
T < T_d \;\; ,\;\; e_\calr \neq 0 \; .
\eeq
Above $T_d$, all traces develop expectation values,
\beq
\left\langle \tr \boldl_\calr \right\rangle \neq 0 \;\; , \;\;
T > T_d \;\; , \;\; \forall \; e_\calr \; .
\eeq
(Implicitly, we always assume that symmetry breaking occurs when
a background
$Z(N)$ field is applied, and then allowed to vanish, in the
appropriate infinite volume limit.)

{\it A priori}, it is not obvious how the traces of $Z(N)$ neutral 
Wilson lines behave in the confined phase, $T < T_d$.
Certainly, they must be nonzero
at {\it all} temperatures.  Even for 
$T < T_d$, they are not protected by the $Z(N)$ symmetry, and so 
are induced by quantum fluctuations at some level.
For three colors, though, numerically we find that 
the expectation value of the renormalized 
octet loop is very small below $T_d$, (\ref{small_neutral}).
This is natural in a matrix model, as discussed at the end of Sec. (IV.C).

There is a counterpart to this
in the behavior of large adjoint Wilson loops.
At zero temperature, a fundamental Wilson loop 
forms a string, with its expectation value the exponential
of the string tension times the area.
Adjoint Wilson loops screen, 
so the adjoint string tension vanishes.  
Greensite and Halpern \cite{largeN} show that at large $N$,
the adjoint string breaks over distances which grow 
as $\sim \log(N)$.  Similarly, the lattice finds that 
the adjoint string only breaks over large distances \cite{lat_sb}.

\subsection{From traces of Wilson lines to Polyakov loops}

Traces of Wilson lines grow with the dimensionality of the
representation.  It is convenient to introduce a normalized
quantity, which approaches one in the obvious perturbative
limit.  From the expectation value of the Wilson line
in a given representation, we 
define the expectation value of the Polyakov loop, $\ell_\calr$, as
\begin{equation}
\left\langle \ell_\calr \right\rangle =
\frac{1}{d_\calr} \; 
\left\langle \tr \boldl_\calr \right\rangle \; .
\label{def_loop}
\end{equation}

The phase of the expectation of a given loop is fixed
by the $Z(N)$ symmetry.
In the perturbative limit at high temperature,
$\boldl_\calr \rightarrow e^{i e_\calr \phi} {\bold 1}_\calr$,
and $\ell_\calr \rightarrow e^{i e_\calr \phi}$.

Using normalized loops, instead of just traces, is most
useful in considering the limit of a large number of colors.
For example, the adjoint loop is
\beq
\elladj = \frac{1}{N^2 - 1} \tr \boldladj
= \frac{N^2}{N^2-1}\left( |\ellfund|^2 - \frac{1}{N^2} \right) \; .
\label{adjoint_loop}
\eeq

The large $N$ limit of this expression cannot be taken directly,
since it involves an operator, $|\ellfund|^2$, and a pure
number.  A large $N$ limit can be taken by comparing expectation values.
This is especially easy at large $N$, as then 
disconnected diagrams dominate,
and all expectation values factorize
\cite{largeN,eguchi_kawai,gocksch_neri,gross_witten,largeN_trick,mixed_action,mixed_action_2,kogut,matrix_deconfining,green_karsch,leutwyler_smilga,matrix_review,gross_taylor}.
For instance, 
\beq
\langle |\ellfund|^2 \rangle = |\langle \ellfund \rangle|^2 \;\; , \;\;
N=\infty \; .
\label{ell_sq_fact}
\eeq
Using this, at large $N$
the expectation value of the adjoint loop is, in magnitude,
identically the square of that for the fundamental loop,
\beq
\langle \elladj \rangle \approx |\langle\ellfund \rangle|^2 + 
 O\left(\frac{1}{N^2}\right) \; . 
\label{large_N_rel_adjoint}
\eeq
As it stands, this is a relationship between bare loops; we
shall see, however, that it survives renormalization.

For the two index tensor representations,
$$
\elltwopm = \frac{2}{N(N \pm 1)} \tr \boldltwopm
$$
\beq
= \frac{N}{N\pm1}
\left( \ellfund^2 \pm \frac{1}{N^2} \tr \boldl_{{ N}}^2 \right)
\label{sym_anti_largeA}
\eeq
Hence at $N=\infty$ the expectation values of the two-index tensor loops are
the square of the fundamental loop,
\beq
\left\langle \elltwopm \right\rangle \approx
\left\langle \ellfund\right\rangle^2 + O\left(\frac{1}{N}\right) \; .
\label{sym_anti_largeB}
\eeq
Notice that the term at infinite $N$ is the same for both
the symmetric and the anti-symmetric representation; the difference
only shows up in corrections in $1/N$.  This generalizes to
higher representations at large $N$.

For the two index tensor
representations, corrections in $1/N$ start with the expectation
value of the operator
\beq
\frac{1}{N} \left( \frac{1}{N} \tr \boldl_{{ N}}^2 \right) \; .
\label{two_index_corrs}
\eeq
In the deconfined phase, we consider the trace of any power of
the Wilson line to be a number of order one, so the trace
of $\boldl_{{ N}}^2$ is like that of
$\boldl_{{ N}}$, a number of order $N$.  Overall, then, this
operator is $\sim 1/N$.  
It is not surprising that corrections to the two-index
loop are $\sim 1/N$, larger than the $\sim 1/N^2$ for the
corrections to the adjoint loop.
These corrections for the two-index loop arise
because the operator has $Z(N)$ charge two, and 
so mix with the operator in (\ref{two_index_corrs}).  
The adjoint loop is 
$Z(N)$ neutral, and so can't mix with this operator.
The analogous operator for the adjoint loop is
$\tr \boldlN^\dagger \boldlN$, which by the unitary
relation of (\ref{unitary_relation}) is a constant.

The generalization to the $\bf N \bf \wedge\bf p$ representations of
the large $N$ limit is immediate,
\beq
\left\langle \ell_{\bf N \wedge \bf p} \right\rangle
=
\left\langle \ellN \right\rangle^{p_+} \;
\left\langle \ellN^* \right\rangle^{p_-} + O\left(\frac{1}{N}\right) \; ,
\label{largeN_wedge}
\eeq
which is (\ref{mean_fieldA}) and (\ref{mean_fieldB}).  
One advantage of using loops is that it is easy to
check overall normalization: up to $Z(N)$ phases, both sides
approach unity in the perturbative limit.

A systematic expansion in $1/N$ proceeds by including operators such as 
\beq
\frac{1}{N^{q-1}} \left( \frac{1}{N} \tr \boldl_{{ N}}^{q} \right) \; ,
\eeq
for integral $q$.
This operator is $\sim 1/N^{q-1}$ in the deconfined phase.  
As $q$ grows, the number of such operators does as well.  We do
not concern ourselves here with the development of a systematic 
expansion, and consider only the leading corrections in $1/N$.
For three colors, our numerical simulations indicate that this
might not be such a bad approximation.  

Even to leading order at large $N$,
some representations involve loops other than the fundamental.
Consider a test baryon, composed of $N$ fundamentals.  Because
of Mandelstam's constraint that 
$\boldl_{{ N}}$ is an $SU(N)$ matrix,
$\det \boldl_{{ N}} = 1$, the term with $N$ powers
of the fundamental loop is part of the identity representation.
Hence at large $N$, the loop for a test baryon behaves as
\beq
\left \langle \ell_{test-baryon} \right \rangle
= \left\langle \ellfund\right\rangle^{N-1} 
\left\langle \frac{1}{N} \; \tr \boldl_{{ N}}^2 
\right\rangle  + O\left(\frac{1}{N}\right)\; .
\eeq
This is illustrated by the decuplet
loop for three colors, (\ref{decup_two}).  

\subsection{Renormalization of Polyakov loops}

With this lengthy introduction aside, we turn to the 
problem at hand, the renormalization of Polyakov loops.  

Remember how mass renormalization usually works, as for
a scalar field, $\phi$, in four spacetime dimensions \cite{zinn}.
If the mass of the field is $m$, and its coupling
$\lambda \phi^4$, 
to one loop order the mass squared receives contributions
\beq
\sim \lambda \int^\Lambda \frac{d^4 k}{k^2 + m^2} \sim
\lambda 
\Lambda^2 \; , \;
\lambda \, m^2 \, \log\left(\frac{\Lambda}{m}\right) \; ;
\label{scalar}
\eeq
a momentum cutoff $\Lambda$ is used to regularize the integral.
The structure at one loop order is generic to perturbation theory:
there are two mass divergences, one proportional to a power of the
cutoff, $\sim \Lambda^2$, and the other, to a logarithm of the cutoff,
$\sim m^2 \log(\Lambda/m)$.  The power divergence
is an additive shift in the bare mass, and 
for a scalar field is inconsequential:
the parameters are tuned to be near a critical point, where the
renormalized mass vanishes.  On the other hand,
the logarithmic divergence is physical, related to
the anomalous dimension for the mass operator.  
A renormalization condition is required to 
fix the value of the renormalized mass at a given scale.

Polyakov loops correspond to a test particle whose mass is taken
to infinity, so their worldline is a straight line.  
This freezes out fluctuations in the time-like direction.
This is obvious in perturbation theory: 
as $\int A_0(\vec{x},\tau) d\tau$ always enters,
only modes which are constant in $\tau$ appear.
Thus the mass divergence of a Polyakov loop
in four spacetime dimensions is like that of a propagating particle
in one less dimension, which is three.  
Similarly, the mass divergences of a scalar field
in four spacetime dimensions, (\ref{scalar}),
are like those of Polyakov loops in five spacetime dimensions.

The ultraviolet divergences of a Wilson line depend only upon
the representation, and not upon (smooth) details of the path.
For the time being, let $\boldl_\calr$ 
denote any Wilson line in a representation $\calr$; we only
assume that the path forms a loop, so that traces of
$\boldl_\calr$ are gauge invariant.
The expectation value of the Wilson line has a mass divergence
which depends upon the length of the loop, ${\cal L}$, as 
\beq
\left\langle \tr \boldl_\calr \right\rangle \; \sim \;
\exp\left( - m^\sdiv_\calr {\cal L} \right) \; .
\eeq

The exponentiation of mass divergences follows from
the analysis of Gervais and Neveu \cite{gervais_neveu}.  
Similar to (\ref{path_integral}), 
they rewrite the Wilson line as a propagator for a fermion
which lives in one dimension, along the path of the loop.
With lattice regularization, the additive mass shift which the
Wilson loop undergoes, $m^\sdiv_\calr$, is no different from
that which (non-gauged) propagating fields also experience,
such as for massive quarks, or scalar fields, $\phi$. 

On the lattice, the exponentiation of mass divergences has
been shown explicitly by Curci, Menotti, and Paffuti 
to $\sim g^4$\cite{ren_loop_old}.

To develop insight into the divergent masses, we compute to one loop order.
In four spacetime dimensions,
\beq
m^\sdiv_\calr 
\sim + \; {\cal C}_\calr g^2 \;  \int^{1/a} \frac{d^3k}{k^2} \sim 
+ \; \frac{{\cal C}_\calr g^2}{a} \; .
\label{divergenceb}
\eeq
We have used a lattice, with lattice spacing $a$, to regularize the theory.  
The exact coefficient of $1/a$ in $m^\sdiv_\calr$
depends on the details of the lattice
discretization, but it is a positive, nonzero number of order one.
In four dimensions, $a m^\sdiv_\calr$ is a
power series in the coupling constant.

For a straight Polyakov loop in four dimensions, this is the only divergence:
there is no anomalous dimension for the corresponding mass.
This is clear to any order in perturbation theory, and occurs
because the mass divergence is like that of a particle which
propagates in three, instead of four, dimensions.

\begin{figure}
\begin{center}
\epsfxsize=.48\textwidth
\epsfbox{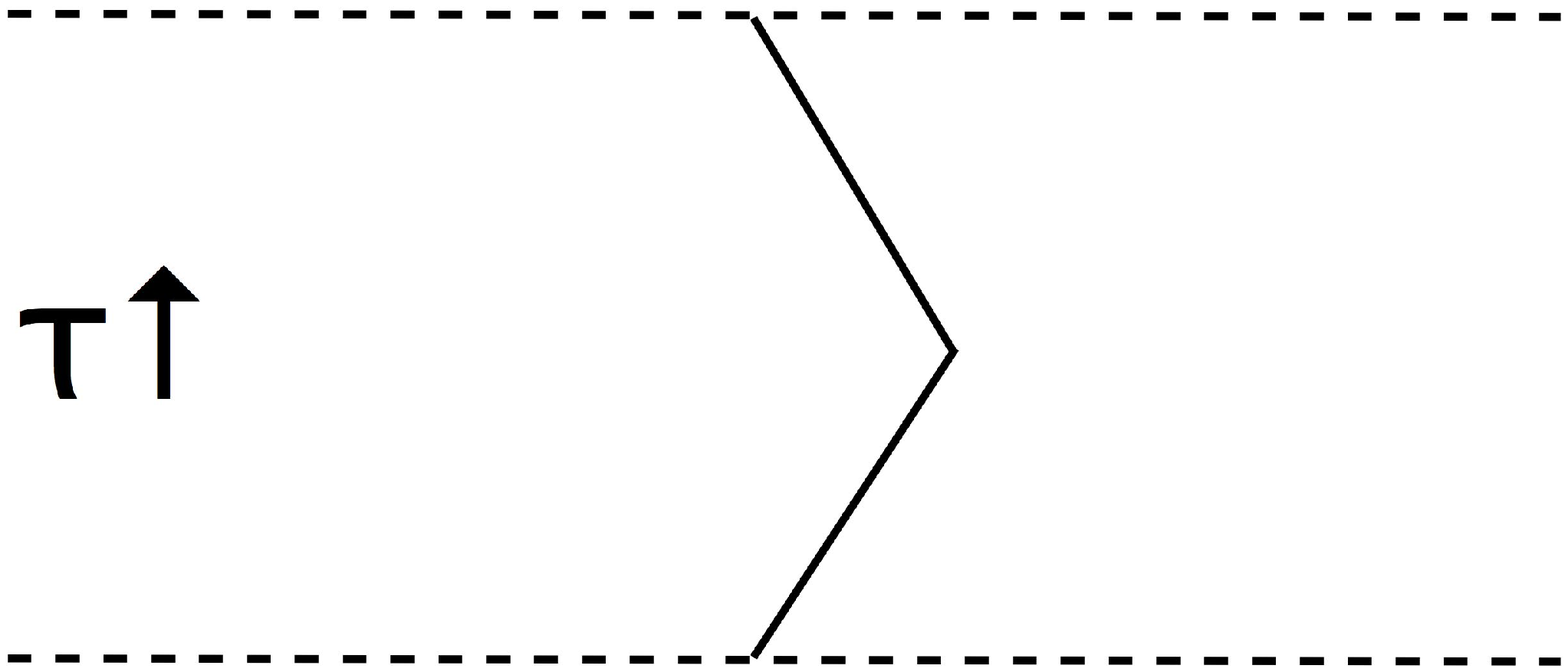}
\end{center}
\caption{A Polyakov loop with two cusps, at 
$\tau = 0 $ and $\tau = 1/(2T)$.  The dotted lines
denote $\tau=0$ and $1/T$.}
\label{fig:cusp}
\end{figure}

Loops can also have cusps
\cite{polyakov_ren,ren_other_a,ren_other_b,ren_cusps,ren_cusp_brem}.  
In order to be periodic in imaginary time, 
the simplest example of a Polyakov loop with cusps has not one,
but two, cusps.
This is illustrated in Fig.~\ref{fig:cusp}, with cusps at $\tau = 0$
and $\tau = 1/(2T)$.  These cusps reflect external probes which deflect
the test particle at these points.
As Polyakov loops, the expectation values of single
loops with cusps only have nontrivial expectation values
in the deconfined phase. 

A cusp generates a logarithmic singularity in four spacetime dimensions
\cite{polyakov_ren,ren_other_a,ren_other_b,ren_cusps,ren_cusp_brem}.
This is not proportional to the length, and so does not contribute
to the divergent mass.  A condition to fix the value of
a renormalized loop with a cusp must be supplied, but this is
standard.  For example, in QCD
loops with cusps are related to the Isgur-Wise function
\cite{ren_cusp_brem}.

It is also interesting to consider loops in three, instead of four,
spacetime dimensions \cite{two_dim}.  
The linear divergence of (\ref{divergenceb}) is now
logarithmic \cite{gervais_neveu},
\beq
m^{div}_\calr \sim + \; {\cal C}_\calr g^2
\int^{1/a} \,  \frac{d^2k}{k^2} \sim 
+ \; {\cal C}_\calr g^2 \log\left(\frac{1}{a}\right) \; ;
\label{three_dims}
\eeq
in three dimensions, the coupling constant $g^2$ has dimensions of mass.
As the divergent mass depends logarithmically upon the lattice
spacing, a condition to fix the value of the renormalized loop
must be supplied \cite{coming2}.
Loops with cusps do not
have new ultraviolet divergences in three spacetime dimensions, although they
do have power like infrared divergences.

Defining wave function renormalization for a Wilson line of
length $\cal L$ as
\beq
\calz_\calr = \exp\left( - \; m^\sdiv_\calr {\cal L} \right) \; ,
\label{define_mass}
\eeq
then the renormalized Wilson line is given by
\beq
\widetilde{\boldl}_\calr \; = \;
\frac{1}{ \calz_\calr} \; \boldl_\calr  \; \; ;
\label{ren_line}
\eeq
as illustrated by the renormalization of Polyakov loops, (\ref{rendloop}).
In the space of all $SU(N)$ invariant tensors, the set of
irreducible representations
form a complete and orthonormal basis \cite{group}.
As this basis is orthonormal, Wilson lines in different
representations do not mix.  Consequently, in different representations
the divergent masses, $m^\sdiv_\calr$, and 
so the renormalization constants $\calz_\calr$,
are independent quantities.

We discuss in the next section how to extract the divergent masses
from lattice simulations using a straight Wilson line.  
In an Appendix we also discuss how the Wilson line might be modified
to alter the mass divergence.

It is illuminating to compute the renormalized loops to one loop
order.  Then it is easiest using dimensional regularization, as then
the divergent mass automatically vanishes.
Following Gava and Jengo \cite{gava,cargese},
the leading correction arises after the Debye mass,
$m_D^2 = N g^2 T^2/3$, is included by resummation.
To lowest order, the correction to the renormalized
loop is 
\beq
\langle \widetilde{\ell}_\calr \rangle - 1 \; \approx \;
- \; {\cal C}_\calr g^2 \int \frac{d^{3-\epsilon} k}{k^2 + m^2_D} 
\; \sim \; + \; {\cal C}_\calr g^2 m_D
\eeq
so that
\beq
\langle \widetilde{\ell}_\calr \rangle 
\approx  \; 1 \; + \;
\frac{\calc_\calr (g^2 N)^{3/2}}{8 \pi N \sqrt{3}} 
+ \; O\left(g^4\right) \; .
\label{pert_loop}
\eeq
(In three spacetime dimensions, 
$\langle \widetilde{\ell}_\calr\rangle - 1 \approx 
\calc_\calr (g^2/T) \log(T/g^2)$ \cite{coming2}.)

In four spacetime dimensions, the leading correction to the
renormalized loop is positive.  Thus in the limit of high
temperature, the loop approaches one from above, and not from
below.  At first sight, this seems paradoxical.
Bare Polyakov loops are traces of $SU(N)$ matrices, and so satisfy a strict
inequality, $|\ell_\calr | \leq 1$.  
For example, on the lattice this holds configuration by configuration.
Instead, renormalized loops satisfy the renormalized constraint,
\beq
|\widetilde{\ell}_\calr| \leq \frac{1}{\calz_\calr} \; .
\label{constraint}
\eeq
Numerically, we find that the divergent masses are uniformly positive, 
\beq
a \, m^\sdiv_\calr > 0 \; ,
\label{positive_mass}
\eeq
for all representations, at all temperatures.  
If so, then the renormalization constant
$\calz_\calr$ always vanishes in the continuum limit,
and there is no constraint on the renormalized loop.
This condition is most natural: otherwise, $\calz_\calr$
diverges as $a\rightarrow 0$, so the renormalized loop must vanish.

The renormalization of a constraint is also familiar
from the non-linear sigma model
in two spacetime dimensions \cite{zinn}.  
If the sigma field is an $SU(N)$ matrix,
then like the bare Wilson line, the bare field is
a unitary matrix.  Because of wave-function renormalization,
however, the renormalized sigma field satisfies a renormalized,
and not a bare, constraint.

In the large $N$ limit, factorization holds.  
This implies that 
\beq
a \, m^\sdiv_\calr \approx (p_+ + p_-) \; a \, m^\sdiv_N \; .
\label{large_N_fact}
\eeq
This is automatic to lowest order in perturbation theory,
where $a m^\sdiv_\calr \sim C_\calr$, remembering that the Casimirs
satisfy (\ref{casimir}).  This also ensures that the
perturbative expression for the
renormalized Polyakov loop, (\ref{pert_loop}), is well behaved
at large $N$.

McLerran and Svetitsky \cite{mclerran}
used the expectation value of a loop
to define the free energy of a test quark.
If this test free energy is defined from the renormalized loop as
$
{\cal F}_\calr
= -\, T \log( |\langle \widetilde{\ell}_\calr \rangle | )
$,
then while it is positive
near $T_d$, from (\ref{pert_loop}) it is negative at high
temperature, 
${\cal F}_\calr \sim - {\cal C}_\calr T/\log(T)^{3/2}$.

While the divergent masses depend upon the
ultraviolet cutoff in a unremarkable manner, the 
renormalization constants are not like those of local
operators.  In four spacetime dimensions, the renormalization
constants of local operators are independent of temperature.
In contrast, the renormalization constants of Polyakov loops
are temperature dependent, but just because the length of
the path for a Polyakov loop is $1/T$.

Renormalization implies that
the only measurable quantities are single traces of Wilson lines.
Consider the most general, gauge invariant combination
of bare Wilson lines possible. For example, start with
$\tr \boldl^{q_1^+}_{\calr^+_1}$, which represents the
propagator for a test quark, in the representation
$\calr^+_1$,
$q_1^+$ times around some fixed loop in spacetime.  Generically,
we can take powers of this trace, and then multiply different
powers of different traces together.  We can also do the same
with conjugate operators.  
By the character expansion \cite{group},
any such combination can be reduced to a linear sum over 
traces of single Wilson lines in different,
irreducible representations:
$$
\left(\tr \boldl^{q_1^+}_{\calr_1^+} \right)^{n_1^+} 
\left( \tr \boldl^{q_2^+}_{\calr_2^+} \right)^{n_2^+} \ldots
\left( \tr (\boldl^\dagger)^{q_1^-}_{\calr_1^-} 
\right)^{n_1^-} \ldots
$$
\beq
= \sum_{\calr} c_{\calr} \; \ell_{\calr} \; 
= \; \sum_{\calr} c_{\calr} \; \calz_\calr \;
\widetilde{\ell}_{\calr} \; .
\label{character}
\eeq
Here all $n$'s and $q$'s are positive integers; the constants
$c_\calr$, and the representations
$\calr$ which one must sum over, are determined by 
group theory \cite{group}. 
Because irreducible representations
form a complete basis over all $SU(N)$ representations \cite{group},
we can insist that only linear powers of Wilson lines appear on the
right hand side.  With a linear sum, renormalization is then just a matter
of replacing bare by renormalized loops.

Assuming that all $m^\sdiv_\calr > 0$, (\ref{positive_mass}),
so the $\calz_\calr$ all vanish as $a\rightarrow 0$,
in the continuum limit 
only the identity representation survives.
%
%
This is of no physical consequence: the physical
quantities, the traces of renormalized Wilson lines, are hidden
in the corrections to this relation, which are 
exponential in $1/a$ as $a\rightarrow 0$.  

This was discovered numerically.
To a high accuracy, we found that
$\langle |\ell_{ 3}|^2 \rangle \approx 1/9$;
corrections varied from $\sim 7\%$ for $N_t = 4$,
to $\sim .2\%$ for $N_t = 10$.  
This is because from (\ref{adjb}), 
$\langle |\ell_{ 3}|^2 \rangle - 1/9 \approx
8 \calz_8 \langle \widetilde{\ell}_{ 8} \rangle /9$;
because of the octet renormalization constant, $\calz_8$,
this is a small quantity.

Previous work on the renormalization of loops at zero temperature
concentrated on loops in the fundamental and adjoint representations,
especially on the case of loops with cusps \cite{ren_cusps,ren_cusp_brem}.
The case of traces of lines which wrap around the same loop
several times, or products of such traces, was neglected.
At nonzero temperature, though,
the natural loops to consider are those at the same point in
space, wrapping around in imaginary time in all possible
ways.  As discussed,
this is equivalent to the set of loops
which wrap around in imaginary time just once, although
in arbitrary representations.

\section{Lattice measurements of $SU(3)$ Polyakov loops}

\def\ignore#1{}
\subsection{General Method}

We turn to the case of three colors.  
Group theory tells us how bare loops are related, 
through expressions such as (\ref{adjoint_loop}).
After renormalization, we do not know how renormalized loops
are related.  Except at very high temperature,
where we can use perturbation theory, (\ref{pert_loop}),
the only way to compute renormalized loops is through
numerical simulations on the lattice.

In this section we discuss how we extract
renormalized Polyakov loops for the lowest, nontrivial representations
of $SU(3)$ color.  
Consider a lattice with $N_s$ steps in each of the three spatial directions,
and $N_t$ steps in the time direction.  At lattice
spacing $a$, the physical temperature $T = 1/(a N_t)$.
As discussed in the Introduction, to extract the mass divergence,
we consider a series of lattices, all at the same
temperature, but with different values of the lattice spacing, $a$, and so
$N_t$; $N_s/N_t$ is kept in fixed ratio.
At a given value of $T/T_d$, we assume that
the logarithm of the
expectation value of a single, bare Polyakov loop can be written
as a power series in $1/N_t$:
\beq
- \;  \log \left \langle \ell_\calr \right\rangle =
f^\sdiv_\calr N_t + f^{ren}_\calr 
+ f^{lat}_\calr \frac{1}{N_t} \; .
\label{divergencea}
\eeq
In four spacetime dimensions, $f^\sdiv_\calr = a m^\sdiv_\calr$.
(In three dimensions, the series is
$f^\sdiv_\calr \log(N_t) + f^{ren}_\calr + f^{lat}_\calr/N_t$.)

Each of the $f_\calr$'s is a power series in the
coupling constant, $g^2$.  On the lattice, this is a series
in the bare coupling constant, and becomes, in the continuum limit,
a series in the renormalized coupling constant.  As such,
the $f_\calr$'s are functions only of the temperature divided by
the renormalization mass scale; or equivalently, of $T/T_d$.
By comparing expectation values at the same temperature, 
but different values of $N_t$, we can extract
$a m^\sdiv_\calr$.  
What remains is the renormalized loop in the continuum limit,
\beq
\left\langle\tildl_\calr \right\rangle
= \exp(- f^{ren}_\calr) \; .
\label{continuumloop}
\eeq
There are also corrections at finite lattice
spacing, $f_\calr^{lat}$.  Near the continuum limit, these
effects begin at $1/N_t$, with
$
f^{lat}_\calr = \sum^{\infty}_{j=1} c_j/N_t^{j-1} 
$.
In weak coupling, $c_1 \sim g^4$; these are corrections
to the one loop term on the lattice, after resumming the
Debye mass.  Corrections in $c_2 \sim g^5$ presumably
arise as lattice corrections to the continuum term, (\ref{continuumloop}).

As is common on the lattice, we work at a fixed ratio of
$N_s/N_t$.  Thus we implicitly assume that the dependence
upon this ratio is negligible in the infinite volume limit.
This can be studied analytically, but requires a careful
treatment of the constant modes.  A perturbative
study is given by Heller and Karsch, especially
Sec. (4.4) \cite{ren_loop_old}.  For now,
we defer this question for future study \cite{coming1}.

\subsection{Lattice Results}

In practice, our method is not quite so trivial.  The difficulty
is that if we require the comparison lattices to have the same
physical temperature, but different $N_t$, then the
deconfining transition occurs at different values of the lattice coupling
constant.  This significantly complicates the analysis.

In the simulations, the Wilson lattice action was used, with
lattice coupling constant $\beta = 6/g^2$.
The number of time steps taken were
$N_t=4$, $6$, $8$, and $10$.
The number of steps in the spatial
direction, $N_s$, was always kept fixed at
$N_s = 3 N_t$; we did not study what happens as this
ratio is varied.
The value of the coupling constant at which the
deconfining transition occurs, $\beta_d$, was
determined by monitoring
the peak in the susceptibility of the triplet loop, to give
the values in Table~\ref{tab:beta_c-vs-Nt}.
\begin{table}[h]
\caption{The lattice coupling constant 
for the deconfining transition, $\beta_d$, at different time steps.}
\label{tab:beta_c-vs-Nt}
\renewcommand{\arraystretch}{1.2}
\renewcommand{\tabcolsep}{1.5pc}
\begin{tabular}{rl}
\hline
$N_t$ & $\beta_d$ \\
\hline
4 & 5.690(5)\\
6 & 5.89(1)\\
8 & 6.055(6)\\
10 & 6.201(5)\\
\hline
\end{tabular}\\[2pt]
\end{table}

\begin{figure}
\begin{center}
\epsfxsize=.48\textwidth
\epsfbox{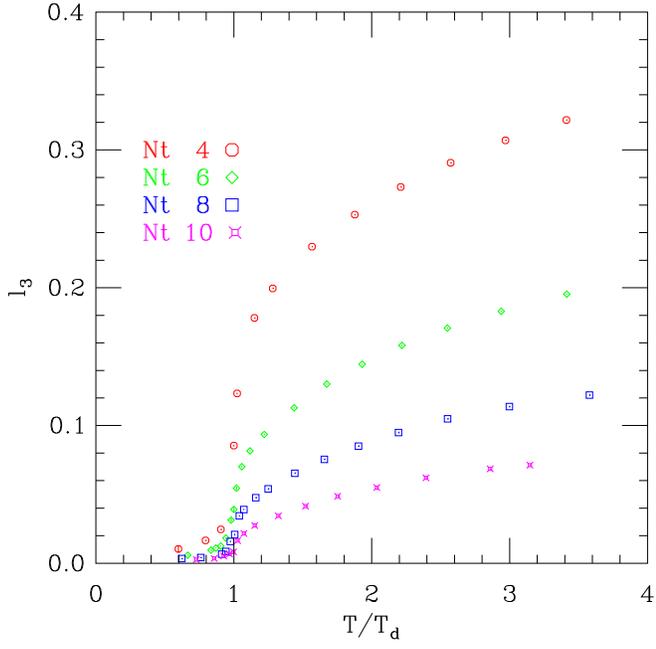}
\end{center}
\caption{The bare triplet Polyakov loop as a function of temperature.}
\label{fig:bare-3}
\end{figure}

\begin{figure}
\begin{center}
\epsfxsize=.48\textwidth
\epsfbox{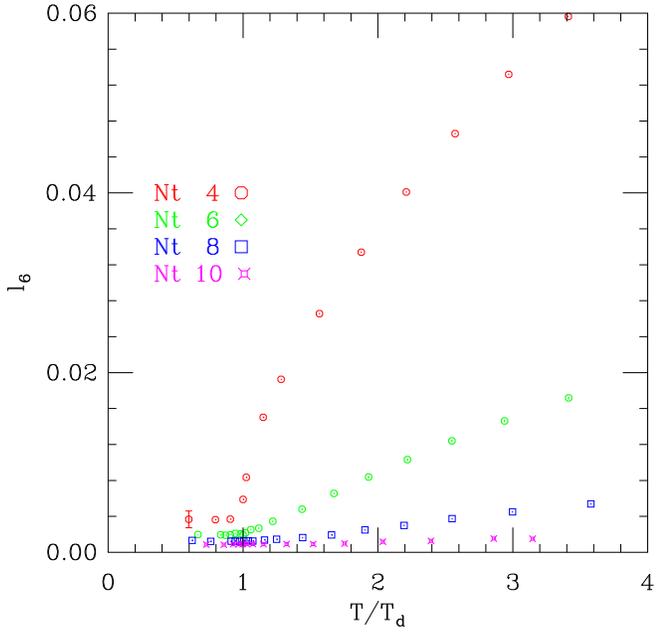}
\end{center}
\caption{The bare sextet Polyakov loop as a function of temperature.}
\label{fig:bare-6}
\end{figure}

\begin{figure}
\begin{center}
\epsfxsize=.48\textwidth
\epsfbox{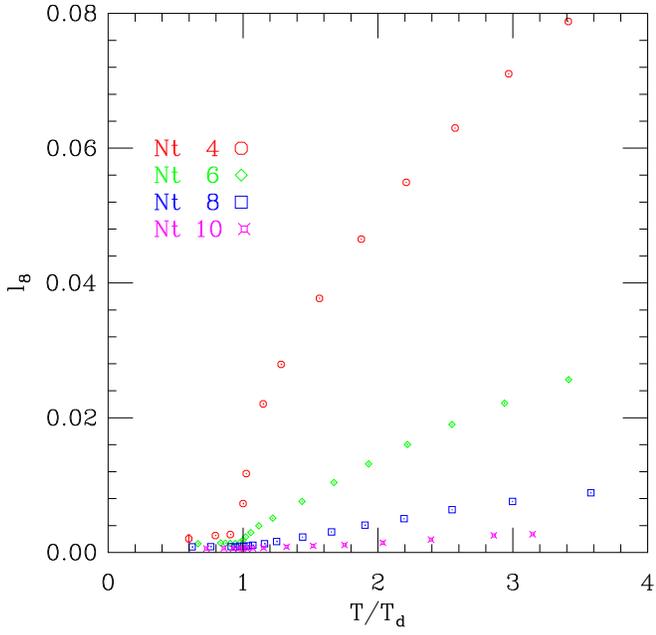}
\end{center}
\caption{The bare adjoint Polyakov loop as a function of temperature.}
\label{fig:bare-8}
\end{figure}

\begin{figure}
\begin{center}
\epsfxsize=.48\textwidth
\epsfbox{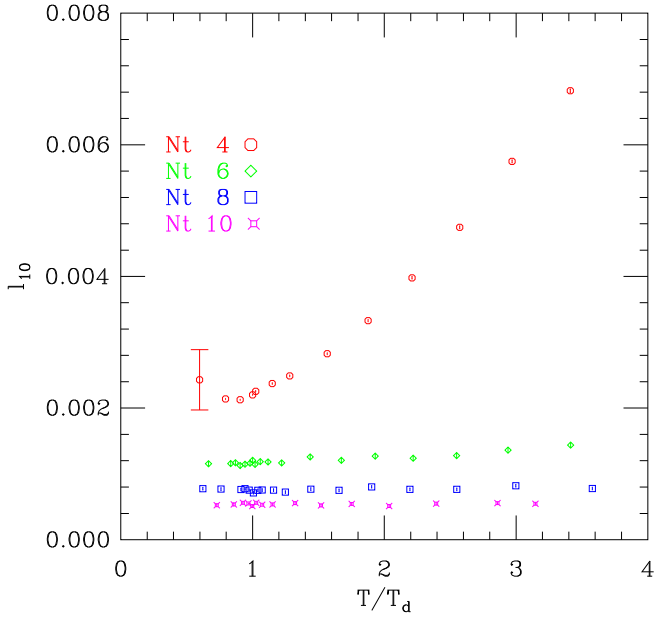}
\end{center}
\caption{The bare decuplet Polyakov loop as a function of temperature.}
\label{fig:bare-10}
\end{figure}

By using non-perturbative renormalization ~\cite{sommer},
the relationship between $\beta$ and the temperature was found to be:
\begin{eqnarray}  
\log \frac{T}{T_d} &=& 1.7139\, (\overline\beta - \overline\beta_d)  - 
                       0.8155\, (\overline\beta^2 -\overline\beta_d^2)\nonumber\\  
		  &+&  0.6667\, (\overline\beta^3 -\overline\beta_d^3)  \; ,
\label{nonpert}
\end{eqnarray}
where 
$\overline\beta \equiv\beta-6$ and $\overline\beta_d \equiv\beta_d-6$.
In terms of physical temperature, our lattices varied from 
$\approx .5 T_d$  to $\approx 3 T_d$.
The lattice calculation was done using  the over relaxed Cabibo-Marinari
 pseudo-heatbath algorithm. Each update step contained 4 heatbath
updates and six overrelaxation steps. A measurement was performed
every 10 update steps. In Table~\ref{tab:Stat} we summarize our
statistics in each case.
Our lattice data for the bare Polyakov loops are presented in
Tables~\ref{tab:bare-loops-Nt-4},~\ref{tab:bare-loops-Nt-6},
\ref{tab:bare-loops-Nt-8}, and~\ref{tab:bare-loops-Nt-10}.
They are also plotted in Figures~\ref{fig:bare-3},~\ref{fig:bare-6},
\ref{fig:bare-8}, and~\ref{fig:bare-10}.
\begin{table}[htb]
\caption{.}
\label{tab:Stat}
\renewcommand{\tabcolsep}{1.2pc}
\begin{tabular}{ll}
\hline
$N_t$ & Measurements \\
\hline
  4 & 10000 \\
  6 & 400   \\
  8 & 400   \\
 10 & 400   \\
\hline
\end{tabular}\\[2pt]
\end{table}
\begin{table}[htb]
\caption{Bare Polyakov loops for $N_t=4$.}
\label{tab:bare-loops-Nt-4}
\renewcommand{\tabcolsep}{.2pc}
\begin{tabular}{lllll}
\hline
$\beta$ & $\langle \ell_3 \rangle$ & $\langle \ell_6 \rangle$ & $\langle \ell_8 \rangle$ & $\langle \ell_{10} \rangle$\\
\hline
5.50 &   0.0104(22) &   0.0037(9) &   0.0020(6) &   0.0024(5) \\
5.60 &  0.01659(9) & 0.003636(19) & 0.002522(19) & 0.002139(11) \\
5.65 &  0.02456(14) & 0.003696(19) & 0.002670(21) & 0.002126(11) \\
5.69 &   0.0854(5) &  0.00588(4) &  0.00726(6) & 0.002202(12) \\
5.70 &   0.1233(4) &  0.00834(4) &  0.01172(6) & 0.002254(12) \\
5.75 &  0.17803(17) &  0.01503(4) &  0.02205(6) & 0.002372(12) \\
5.80 &  0.19956(15) &  0.01925(5) &  0.02791(6) & 0.002490(13) \\
5.90 &  0.22978(14) &  0.02657(5) &  0.03772(6) & 0.002825(15) \\
6.00 &  0.25301(13) &  0.03340(5) &  0.04649(6) & 0.003328(17) \\
6.10 &  0.27307(13) &  0.04010(6) &  0.05492(6) & 0.003979(19) \\
6.20 &  0.29075(13) &  0.04660(6) &  0.06298(7) & 0.004747(20) \\
6.30 &  0.30695(12) &  0.05319(6) &  0.07103(7) & 0.005748(22) \\
6.40 &  0.32174(12) &  0.05964(6) &  0.07880(7) & 0.006821(24) \\
\hline
\end{tabular}\\[2pt]
\end{table}
\begin{table}[htb]
\caption{Bare Polyakov loops for $N_t=6$.}
\label{tab:bare-loops-Nt-6}
\renewcommand{\tabcolsep}{.2pc}
\begin{tabular}{lllll}
\hline
$\beta$ & $\langle \ell_3 \rangle$ & $\langle \ell_6 \rangle$ & $\langle \ell_8 \rangle$ & $\langle \ell_{10} \rangle$\\
\hline
5.70 &  0.00592(16) &  0.00197(5) &  0.00131(5) &  0.00115(3) \\
5.80 &  0.00963(26) &  0.00195(5) &  0.00143(6) &  0.00116(3) \\
5.82 &   0.0111(3) &  0.00191(5) &  0.00139(5) &  0.00117(3) \\
5.84 &   0.0125(3) &  0.00195(5) &  0.00135(5) & 0.001126(29) \\
5.86 &   0.0183(5) &  0.00210(5) &  0.00132(5) & 0.001148(28) \\
5.88 &   0.0314(9) &  0.00205(5) &  0.00150(6) &  0.00116(3) \\
5.89 &   0.0391(10) &  0.00195(5) &  0.00183(7) &  0.00121(3) \\
5.90 &   0.0545(9) &  0.00217(6) &  0.00231(8) & 0.001144(29) \\
5.92 &   0.0702(6) &  0.00254(6) &  0.00296(9) &  0.00119(3) \\
5.95 &   0.0816(5) &  0.00268(7) &  0.00397(9) &  0.00118(3) \\
6.00 &   0.0935(5) &  0.00346(8) &  0.00508(10) &  0.00117(3) \\
6.10 &   0.1128(4) &  0.00480(9) &  0.00759(11) &  0.00126(3) \\
6.20 &   0.1301(4) &  0.00654(9) &  0.01038(11) &  0.00120(3) \\
6.30 &   0.1445(4) &  0.00837(10) &  0.01314(11) &  0.00127(3) \\
6.40 &   0.1581(4) &  0.01031(10) &  0.01604(13) &  0.00124(3) \\
6.50 &   0.1707(4) &  0.01241(11) &  0.01901(12) &  0.00128(3) \\
6.60 &   0.1829(4) &  0.01462(12) &  0.02216(14) &  0.00136(4) \\
6.70 &   0.1954(4) &  0.01718(12) &  0.02565(14) &  0.00144(4) \\
\hline
\end{tabular}\\[2pt]
\end{table}
\begin{table}[htb]
\caption{Bare Polyakov loops for $N_t=8$.}
\label{tab:bare-loops-Nt-8}
\renewcommand{\tabcolsep}{.2pc}
\begin{tabular}{lllll}
\hline
$\beta$ & $\langle \ell_3 \rangle$ & $\langle \ell_6 \rangle$ & $\langle \ell_8 \rangle$ & $\langle \ell_{10} \rangle$\\
\hline
5.80 &  0.00339(9) &  0.00133(4) &  0.00083(3) & 0.000775(20) \\
5.90 &  0.00421(11) &  0.00122(3) &  0.00086(3) & 0.000770(20) \\
6.00 &  0.00668(18) &  0.00124(3) &  0.00087(3) & 0.000763(20) \\
6.02 &  0.00858(26) &  0.00126(4) &  0.00082(3) & 0.000775(20) \\
6.04 &   0.0159(5) &  0.00121(3) &  0.00095(4) & 0.000754(19) \\
6.06 &   0.0209(5) &  0.00129(3) &  0.00092(3) & 0.000713(17) \\
6.08 &   0.0345(4) &  0.00129(3) &  0.00099(4) & 0.000751(20) \\
6.10 &   0.0391(3) &  0.00126(4) &  0.00108(4) & 0.000755(20) \\
6.15 &  0.04762(26) &  0.00138(4) &  0.00131(5) & 0.000754(20) \\
6.20 &  0.05400(25) &  0.00146(4) &  0.00164(5) & 0.000724(20) \\
6.30 &  0.06541(24) &  0.00164(5) &  0.00228(6) & 0.000768(21) \\
6.40 &  0.07540(24) &  0.00194(5) &  0.00305(6) & 0.000751(19) \\
6.50 &  0.08501(22) &  0.00249(5) &  0.00407(5) & 0.000804(18) \\
6.60 &  0.09492(24) &  0.00299(5) &  0.00502(6) & 0.000765(19) \\
6.70 &  0.10490(24) &  0.00375(5) &  0.00635(6) & 0.000766(21) \\
6.80 &  0.11377(25) &  0.00450(6) &  0.00757(6) & 0.000819(20) \\
6.90 &  0.12214(24) &  0.00540(5) &  0.00886(7) & 0.000778(20) \\
\hline
\end{tabular}\\[2pt]
\end{table}
\begin{table}[htb]
\caption{Bare Polyakov loops for $N_t=10$.}
\label{tab:bare-loops-Nt-10}
\renewcommand{\tabcolsep}{.2pc}
\begin{tabular}{lllll}
\hline
 $\beta$ & $\langle \ell_3 \rangle$ & $\langle \ell_6 \rangle$ & $\langle \ell_8 \rangle$ & $\langle \ell_{10} \rangle$\\
\hline
6.00 &  0.00272(9) &  0.00088(3) &  0.00058(3) & 0.000525(18) \\
6.10 &  0.00362(13) &  0.00085(3) &  0.00064(4) & 0.000537(22) \\
6.15 &  0.00524(15) & 0.000904(24) & 0.000588(23) & 0.000560(13) \\
6.18 &  0.00672(16) & 0.000906(18) & 0.000615(17) & 0.000550(11) \\
6.20 &  0.00839(27) & 0.000909(23) & 0.000633(24) & 0.000512(14) \\
6.22 &  0.01651(22) & 0.000927(18) & 0.000627(18) & 0.000561(11) \\
6.25 &  0.02167(17) & 0.000929(18) & 0.000641(18) & 0.000532(11) \\
6.30 &  0.02741(22) &  0.00091(3) &  0.00070(4) & 0.000537(18) \\
6.40 &  0.03444(24) &  0.00092(4) &  0.00084(4) & 0.000555(21) \\
6.50 &  0.04142(24) &  0.00092(4) &  0.00099(4) & 0.000521(20) \\
6.60 &  0.04861(18) &  0.00098(3) &  0.00111(4) & 0.000543(17) \\
6.70 &  0.05505(18) &  0.00118(3) &  0.00146(5) & 0.000511(17) \\
6.80 &  0.06204(19) &  0.00128(4) &  0.00193(5) & 0.000549(16) \\
6.90 &  0.06852(19) &  0.00154(3) &  0.00255(5) & 0.000555(16) \\
6.95 &  0.07130(18) &  0.00152(3) &  0.00270(4) & 0.000546(15) \\
\hline
\end{tabular}\\[2pt]
\end{table}

As the relationship between the lattice coupling constant
and the physical temperature, (\ref{nonpert}) is nonlinear,
it is not automatic ensuring that the temperature is the
same when $N_t$ changes.  Thus
we resort to interpolation, measuring the loops 
on a fixed grid, in $\beta$, for each $N_t$. 
For $N_t=4$, $6$, and $8$ we have linearly interpolated the Polyakov
loop values to the $T/T_d$ values 
at which the measurements for $N_t=10$ were done.
Then, for each value of $T/T_d$,
the expectation value of the bare Polyakov loop was fit to
(\ref{divergencea}) --- (\ref{continuumloop}). 

In our measurements,
we see no statistically significant term $\sim 1/N_t$,
$f^{lat}_\calr \approx 0$.  Such terms will
presumably be revealed by more precise measurements.
The success of a fit to (\ref{divergencea}) indicates that on
the lattice, the divergent mass $m^\sdiv_\calr$ does exponentiate.  

We stress that we make no assumptions about any of the
functions $f_\calr$.  
At a given value of the temperature, the logarithm of
the expectation value of the bare loop,
$\log(\langle \ell_\calr \rangle)$, is a power series
in $1/N_t$, beginning as $\sim N_t$,
(\ref{divergencea}) --- (\ref{continuumloop}).
Given the lattice data, for loops at the same physical temperature
and different values of $N_t$, there is nothing left over to adjust.

\begin{figure}
\begin{center}
\epsfxsize=.48\textwidth
\epsfbox{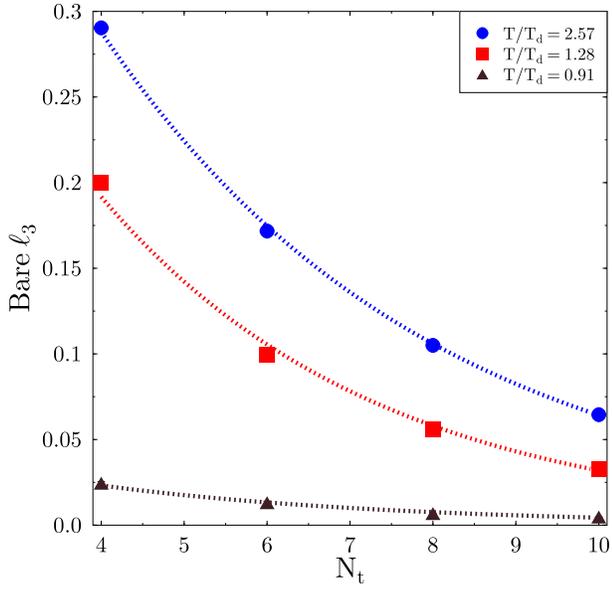}
\end{center}
\caption{The bare 
triplet Polyakov loop at fixed temperature, as a function of $N_t$.}
\label{fig:Fit}
\end{figure}

In Fig.~\ref{fig:Fit} we present a typical fit for the
expectation value of the bare triplet loop.  It is clear that 
the bare loop 
decreases, with increasing $N_t$, for all temperatures measured.  

In Table~\ref{tab:twoTd}, we give the expectation values of the bare
Polyakov loops for different representations.
We chose the smallest $N_t$, $N_t = 4$, where the
signals are greatest.  For reference, we also
include the Casimirs of the different representations.
\begin{table}[h]
\caption{Approximate expectation values of the bare Polyakov loop, and
Casimirs, for $N_t = 4$ and $T = 2 T_d$.}
\label{tab:twoTd}
\renewcommand{\arraystretch}{1.2}
\renewcommand{\tabcolsep}{1.5pc}
\begin{tabular}{rll}
\hline
$\calr$ & $\langle \ell_\calr\rangle$ & ${\cal C}_\calr$ \\
\hline
$ 3$  & $.25$    & $4/3$  \\
$ 8$  & $.04$    & $3$           \\
$ 6$  & $.035$   & $10/3$ \\
$10$ & $.004$   & $6$            \\
\hline
\end{tabular}\\[2pt]
\end{table}

For all representations, the signal decreases with increasing $N_t$.
This indicates that every $a m^\sdiv_\calr$ is positive, as suggested before,
(\ref{positive_mass}).  
We were only able to measure a signal for the decuplet loop on
the smallest lattice, $N_t = 4$.
Perhaps a modified Polyakov loop, as discussed
in the Appendix, might help.  We do not discuss the decuplet loop further.

\begin{figure}
\begin{center}
\epsfxsize=.48\textwidth
\epsfbox{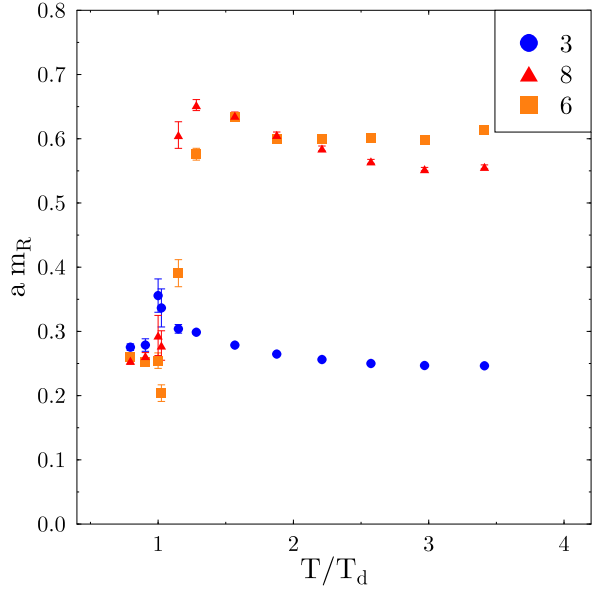}
\end{center}
\caption{The divergent masses, times lattice spacing,
versus temperature.}
\label{fig:mR}
\end{figure}

Fig.~\ref{fig:mR} shows our results for the product of the lattice spacing
times the divergent mass, $a m^\sdiv_R$.
For the triplet loop, 
this product doesn't vary much with temperature, 
but those for the sextet and octet loops do.
At the highest temperature, $\approx 3 T_d$, the values
are approximately
what one expects from lowest order in perturbation theory,
where the divergent masses scale like the Casimir of
the representation, $a m^\sdiv_\calr \sim C_\calr$, (\ref{divergenceb}).
As the temperature decreases below $\approx 1.5 T_d$, though,
the perturbative ordering of $a m^\sdiv_6 > a m^\sdiv_8$ is
reversed, with $a m^\sdiv_8 > a m^\sdiv_6$.  
All divergent masses 
are approximately equal below $T_d$, although
the signals are poor.

\begin{figure}
\begin{center}
\epsfxsize=.48\textwidth
\epsfbox{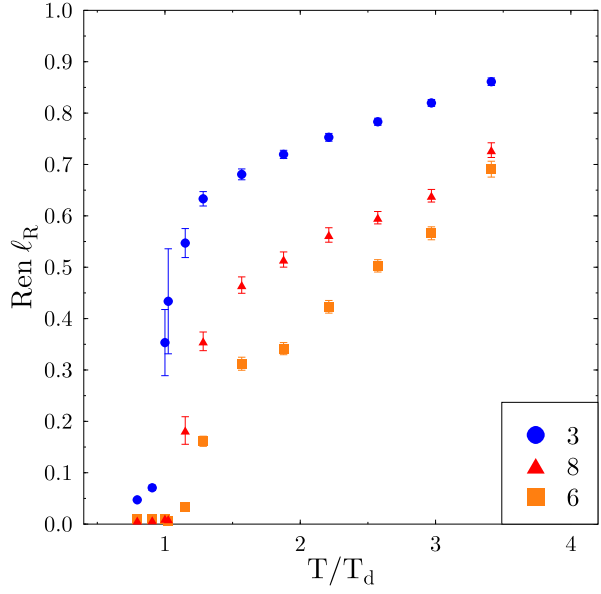}
\end{center}
\caption{Renormalized Polyakov loops as a function of temperature.}
\label{fig:RenPloops}
\end{figure}

After dividing by the renormalization constant, we obtain
the renormalized loops of Fig.~\ref{fig:RenPloops}.
Below $T_d$, the triplet and
sextet fields, which carry $Z(3)$ charge, should vanish in the
infinite volume limit.
What is striking is that below $T_d$, the $Z(3)$ neutral adjoint loop ---
which could be nonzero --- is also too small for us to measure:
\beq
\left\langle \widetilde\ell_8
\right \rangle \approx 0 \;\;\; , \;\;\; T < T_d \; .
\label{small_neutral}
\eeq
Previous studies found this for bare adjoint loops on small lattices,
such as $N_t = 4$ \cite{teper,susy2,damgaard,twocolors}.

In the deconfined phase, the triplet loop is always greatest,
followed by the octet, and then the sextet loop.  
At $T_d^+$, the triplet loop jumps to a relatively
large value, $\approx .4$, although the exact value is
not very well determined.  Due to the increase in correlation
lengths, there is critical slowing down near $T_d$, and much
more careful studies are required.

\begin{figure}
\begin{center}
\epsfxsize=.48\textwidth
\epsfbox{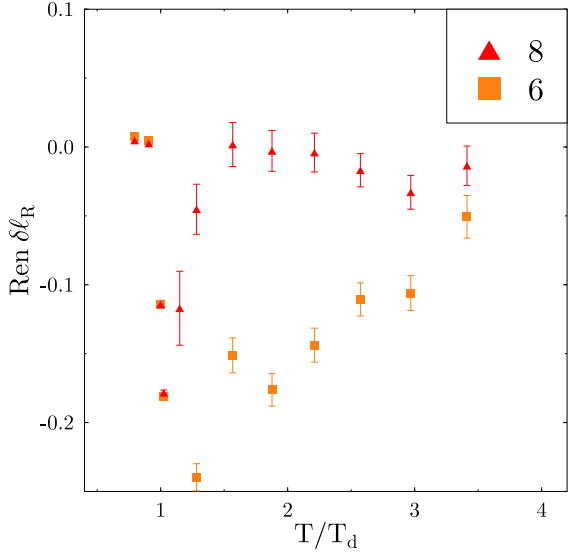}
\end{center}
\caption{Difference loops: renormalized Polyakov loops, minus their
large $N$ values.}
\label{fig:LoopsDiffs}
\end{figure}

We then computed the expectation value of the difference
between the sextet and octet loops, and their large $N$ limit,
(\ref{sextet_diff}) and (\ref{octet_diff}).  
The results are presented
in Fig. (\ref{fig:LoopsDiffs}).
Numerically, we find that both difference loops
are negative.  They each look like a ``spike'' down,
with a maximum near $T_d$.  The sextet spike is
larger, $|\deltilell_{ 6}| \leq .25$, with
a maximum at $\approx 1.25 T_d$.  
The octet spike is smaller, $|\deltilell_{ 8}| \leq .2$, 
with a maximum nearer $T_d$, at $\approx 1.1 T_d$.
The octet spike falls off much more quickly than the sextet
spike with increasing temperature: by $\approx 1.5 T_d$,
the octet difference vanishes, while the sextet difference
persists all of the way to $\approx 3 T_d$.

The most important feature of the difference loops is
their overall magnitude: each is significantly smaller than
one.  As discussed in the Introduction, this indicates that
factorization, which is exact
for $N=\infty$, is approximately correct for $N=3$.
The difference loops will be discussed further in the
next section.

We conclude this section by contrasting our method for
determining renormalized Polyakov loops 
with that of Kaczmarek {\it et al.} \cite{bielefeld_ren}.  
They measure 
$\langle \ell_{\overline{ 3}}(\vx)
\ell_{{ 3}}(0)\rangle - |\langle \ell_{{ 3}}\rangle|^2$, 
and extract $\calz_{{ 3}}$ by
comparing with perturbation theory at short distances.  
Thus
they need to measure two-point functions, although at just one value of $N_t$.
We only need to measure one-point
functions, but must do so at several values of $N_t$.
Our results agree approximately with theirs; near
$T_d$, we both have 
$|\langle \widetilde{\ell}_{ 3} \rangle| \approx .4$.  It disagrees in
that by $3 T_d$, their 
$|\langle \widetilde{\ell}_{ 3} \rangle| \approx 1.0$, 
while ours is $\approx .9$.
We do not know the reason for the differences between our results,
although it could simply be due to the effects of finite
lattice spacing.

\section{Matrix Models for Renormalized Loops}

\subsection{Effective Theories}

We now discuss various effective models for 
the deconfining phase transition 
\cite{sv_yaffe,banks_ukawa,resum,shuryak,susy1,susy2,susy3,susy4,susy6,susy7,susy8,susy9,kinetic,pol_loop_a,pol_loop_b,pol_loop_c,cargese,sannino,pol_loop_misc,gross_witten,largeN_trick,mixed_action,mixed_action_2,kogut,matrix_deconfining,green_karsch,leutwyler_smilga,matrix_review,damgaard},
and fit the renormalized triplet loop to a
simple matrix model \cite{kogut,green_karsch,damgaard}.
We then discuss the Gross--Witten point at large $N$, and how
it may be related to three colors.

Numerical simulations on the lattice
suggest that the transition is of second order
for two colors \cite{twocolors}
and of first order for three \cite{flavor_ind}, 
four \cite{four_colors,teper}, and six \cite{teper} colors.
Lucini, Teper, and Wenger \cite{teper} present evidence
that the latent heat grows $\sim N^2$ for these values of $N$.
{}From this we presume that
the transition is of first order for all $N\geq 3$.  
Arguments for a first order transition at infinite $N$
have also been given by Gocksch and Neri \cite{gocksch_neri}.

We start with an effective lattice theory in the purely
spatial dimensions.  
The actual value of the lattice spacing is irrelevant:
all that matters is that it is much smaller than any physical
length scale.  Thus we concentrate on the region near $T_d$,
and ignore $T \rightarrow \infty$.  

The simplest approach is to follow Svetitsky and Yaffe \cite{sv_yaffe},
and construct an effective theory just for the Polyakov loop
in the fundamental representation.  This is certainly fine if
the transition is of second order, as the only critical
field should be that for the fundamental loop.
For the second order transition of two colors, this predicts
that the universality class is that of the Ising model,
in agreement with the observed critical exponents \cite{twocolors}.
For three colors, because $Z(3)$ symmetry allows a cubic invariant,
this approach also predicts that the transition is invariably of first order.
For more than three colors, though, it is not clear why
the transition should be of first order.  
It would be if the interactions of the
$Z(N)$ spins were like those of a Potts
model, but they are not at high temperature \cite{cargese}.
Loops in higher representations can be introduced into this
model, as new types of $Z(N)$ spins.  However, at large $N$
factorization does not follow naturally, but must be imposed by hand.

A variant of this approach is the Polyakov loop model 
\cite{pol_loop_a,pol_loop_b,pol_loop_c,cargese}.
This starts with a potential for the (renormalized) fundamental loop,
and assumes that the pressure is 
correlated to the value of the potential at its minimum.  We presume
that the same holds for
the potential of the matrix model, secs. IV.B and IV.C.

The closest model to the underlying gauge theory is 
a gauged, nonlinear sigma model for the Wilson line $\boldl_{ N}$
\cite{banks_ukawa,pol_loop_a}.
In such a model, 
Higgs phases, in which the $SU(N)$ symmetry is
spontaneously broken, can appear, but these are not
expected to arise if there are no dynamical scalars about.
Remember that deconfinement is not a Higgs effect:
only $Z(N)$ is broken in the deconfined phase, not $SU(N)$.

Assuming that Wilson lines form the essential degrees of freedom
in the effective theory, we take as the partition function:
\beq
{\cal Z} = \int \Pi \; d\boldlN(i)
\; \exp \left(- {\cal S}(\ell_\calr(i)) \right) \; ;
\label{partition}
\eeq
$i$ denotes lattice sites in the spatial directions.
We assume that only Polyakov loops enter into the effective
action, and consider only constant solutions in mean field theory.
We neglect all kinetic terms, including those for the Wilson line,
Polyakov loops, and color magnetic fields \cite{kinetic}.
The effects of fluctuations, which are controlled by these
kinetic terms, can be important, especially in 
three spacetime dimensions \cite{coming2}.

At each site, 
the measure in (\ref{partition}) implicitly contains a constraint
to enforce that $\boldlN(i)$ is a $SU(N)$ matrix.  As such,
loops $\ell_\calr(i)$ are constructed by the usual
relations of group theory.  This is most convenient, since then 
all loops automatically have the right
$Z(N)$ charge, and satisfy factorization at large $N$.
Moreover, at each site we can use the character expansion, 
(\ref{character}), to reduce any product of loops to a linear
sum.  This vastly restricts the number
of possible couplings which can arise.

In a sigma model over a symmetric space,
the trace of $\boldl_{ N}$ is everywhere some fixed constant \cite{zinn}.
In the present instance, however, the trace
of the Wilson line is not constant.  Thus the
action also includes a potential for the Wilson line \cite{pol_loop_a}.
Requiring the action to be $Z(N)$ invariant, this is 
a sum over Polyakov loops with vanishing $Z(N)$ charge:
\beq
{\cal W} = 
N^2 \; \Sigma_i \; \Sigma_{\calr}^{e_\calr = 0} 
\;\; \gamma_\calr \; \ell_\calr(i) \; ;
\label{zN_potential}
\eeq
the $\gamma_\calr$ are coupling constants.  This
series begins with the adjoint loop.  

Next, loops on one site can interact with those on another site:
\beq
{\cal S}_\calr = - \frac{N^2}{3}\;
\Sigma_{i,\hat{n}} \; \Sigma^{e_\calr + e_{\calr'}=0}_{\calr, \calr'} \;
\beta_{\calr,\calr'} \;
{\rm Re} \; \ell_\calr(i) \ell_{\calr'}(i + \hat{n}) \; .
\label{eff_act_RRp}
\eeq
$\rm Re$ denotes the real part, and $\beta_{\calr,\calr'} =
\beta_{\calr',\calr}$.  We just write nearest neighbor
interactions, given by the sum over the lattice vector $\hat{n}$,
but this is inessential.
The sum over representations is restricted by the requirement that
the total $Z(N)$ charge of each term is zero, modulo $N$.
There are both diagonal couplings, where $\calr'$ is the representation
conjugate to $\calr$, and off-diagonal couplings,
where $\calr' \neq \calr^*$.  This is the complete
set of independent couplings.

\subsection{Mean Field and Matrix Models for $N=3$}

For three colors, the simplest action includes just
the triplet loop \cite{matrix_deconfining}
\beq
{\cal S}_3 = - 3 \beta_3 \;
\Sigma_{i,\hat{n}} \; {\rm Re} \; \ellthree(i) \ellthree^*(i + \hat{n})
\; ;
\label{eff_act_3}
\eeq
$\beta_{3,3^*} \equiv \beta_3$.
We use this to develop a mean field approximation, replacing
all nearest neighbors by an average value.
On a cubic lattice in three dimensions,
if the value of each neighboring spin is
$\ell_0 = \langle \ell_{ 3} \rangle$ 
in (\ref{eff_act_3}), the partition function of
(\ref{partition}) reduces to one for a single site,
\beq
\calz = \int \; d\boldl_{ 3}
\; \exp\left( + 18 \beta_3 \ell_0 {\rm Re} \ell_{ 3} \right) 
\equiv \exp(- 9 {\cal V}) \; .
\label{zss_3}
\eeq
This is a matrix model, but one
whose coupling constant depends on the value of the condensate,
$\ell_0$.  We introduce the single site potential, ${\cal V}$.
The mean field condition is that the average value, computed
with this action, is equal to the assumed value \cite{zinn}:
\beq
\ell_0 = - \frac{1}{2 \beta_3}
\; \frac{\partial}{\partial \ell_0} {\cal V} \; .
\label{mean_field_3}
\eeq

As $\boldlN$ is an $SU(N)$ matrix, 
it is the unitary transformation of a diagonal matrix, with
$N-1$ independent eigenvalues.
When $N=2$, the integral like (\ref{zss_3})
is elementary, and can be evaluated in terms of Bessel 
functions.  These can also be done when $N>2$, but 
we found it easier to simply evaluate it numerically.
Explicitly, with
${\bf L}_{{ 3}} = U\,{\rm diag}(\exp(i\theta_1),\exp(i\theta_2),
\exp(-i(\theta_1+\theta_2))\,U^\dagger$,
the normalized Haar measure, including the van der Monde determinant, is
$$
d\boldl_{ 3} = 
\frac{1}{3\pi^2} \,(1-\cos(\theta_1-\theta_2)) (1-\cos(2\theta_1+\theta_2))
$$
\beq
\,(1-\cos(\theta_1+2\theta_2))\; d\theta_1 d\theta_2\,\, .
\eeq

This mean field theory was studied in the context of the 
deconfinement transition by several groups 
\cite{kogut,matrix_deconfining,green_karsch,damgaard}.
Like the lattice data, the transition is of second order
for $N=2$, and first order when $N\geq3$.
For three colors, Damgaard \cite{damgaard}
used eq.\ (\ref{mean_field_3}) 
to compute the expectation value of the triplet loop;
expectation values for the 
sextet, adjoint and decuplet loops were then computed from that.
Damgaard compared the results of this mean field theory to lattice
data for bare Polyakov loops, with $N_t = 3$,
by Markum, Faber, and Meinhart \cite{ren_loop_old},
finding qualitative agreement.

We stress that the approximate agreement between this mean
field theory, and lattice data at small $N_t$, is in some sense
fortuitous.  For small $N_t$, the $\calz_\calr$ 
are not much different from one,
and so the bare values are not far from the renormalized
values; even so, they are not identical.
To see this in another way,
we computed the ratio of the difference 
loop, to the loop itself, for the bare octet loop:
$|(\langle \ell_8\rangle - |\langle \ell_3\rangle|^2)/\langle \ell_8\rangle|$.
This ratio is $\sim 50\%$
at $N_t=4$, and increases to $\sim 100-200\%$
for $N_t = 10$.  This is to be compared with the values for
the renormalized octet difference loop in fig. (4), which is $\leq 12\%$.
Thus while renormalized loops satisfy factorization, bare loops do not.

In this vein, recently Dittmann, Heinzl, and Wipf 
computed the effective potential for bare doublet loops
in a pure $SU(2)$ gauge theory 
\cite{pol_loop_misc}.  Because only renormalized loops
satisfy factorization, we suggest that the effective potential
for renormalized loops is much simpler than that for bare loops.

\begin{figure}
\begin{center}
\epsfxsize=.48\textwidth
\epsfbox{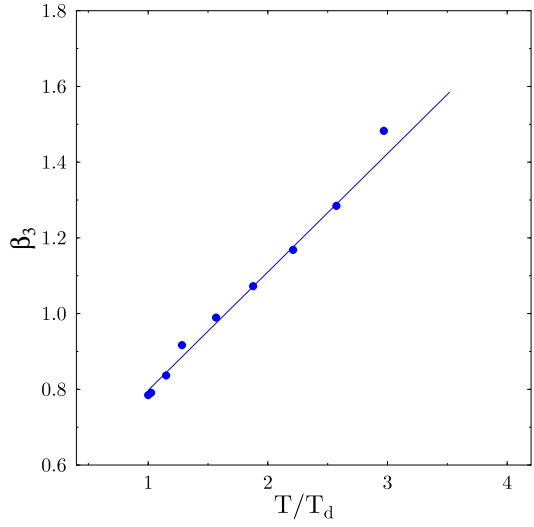}
\end{center}
\caption{The fundamental coupling constant extracted from the lattice 
data.
The circles are the fundamental coupling computed from lattice data for 
the
fundamental loop. The error bars on the extracted points are smaller than 
data points. The smooth line is the linear fit to the extracted fundamental 
coupling with the errors quoted in (\ref{eq:beta3fit}).  }
\label{fig:beta3coup}
\end{figure}

We next compare the solution of
mean field theory, eq.\ (\ref{mean_field_3}), 
to our lattice data for the renormalized triplet loop.
We find that a linear relationship between the
mean field coupling constant, $\beta_3$, and the temperature,
$T/T_d$, is approximately valid.
A least squares fit gives
\beq
\beta_3 = (0.46 \pm 0.02) + (0.33 \pm 0.02) \, \frac{T}{T_d} \,\, .
\label{eq:beta3fit}
\eeq
The computed coupling, as well as the fitted curve, is shown in 
Fig.\ \ref{fig:beta3coup}.  A quadratic term was included 
in the fit but the coefficient was found to be zero within 
the error bars.  Significant deviations are seen at both
the highest temperature, $\approx 3T_d$, and also at the
two points closest to $T_d$; 
see the discussion at the end of this subsection.
The approximate linear relationship between 
$\beta_3$ and the temperature is typical
of mean field theory for spin models \cite{zinn}.  

\begin{figure}
\begin{center}
\epsfxsize=.48\textwidth
\epsfbox{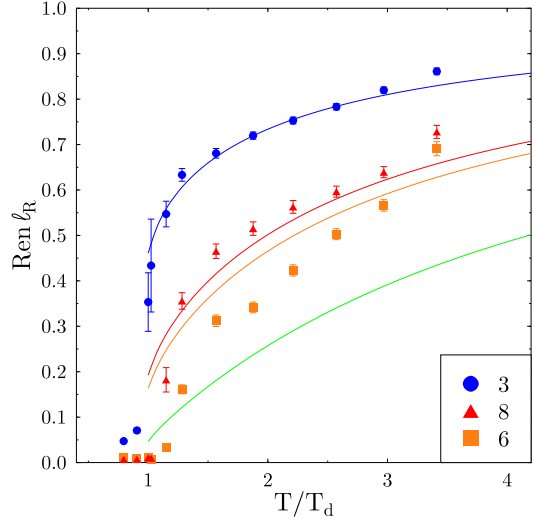}
\end{center}
\caption{The values of the Polyakov loops from the matrix model, using the
linear relationship between the coupling and the temperature.
The lower-most (green) line corresponds to the value of the decuplet
loop, for which there is no lattice data.
}
\label{fig:FundActionConds}
\end{figure}

Using this relationship between the mean field coupling and
the temperature, we then computed mean field results
for the sextet, octet, and decuplet loops.  The comparison
to our lattice data are shown in 
Fig.\ \ref{fig:FundActionConds}.  Notice that although
we could not extract from the lattice a signal for the
decuplet loop, the mean field theory predicts that 
while the decuplet loop is less than the sextet, it
is not that small; for example, $\ell_{10}(3 T_d) \approx .4$.

\begin{figure}
\begin{center}
\epsfxsize=.48\textwidth
\epsfbox{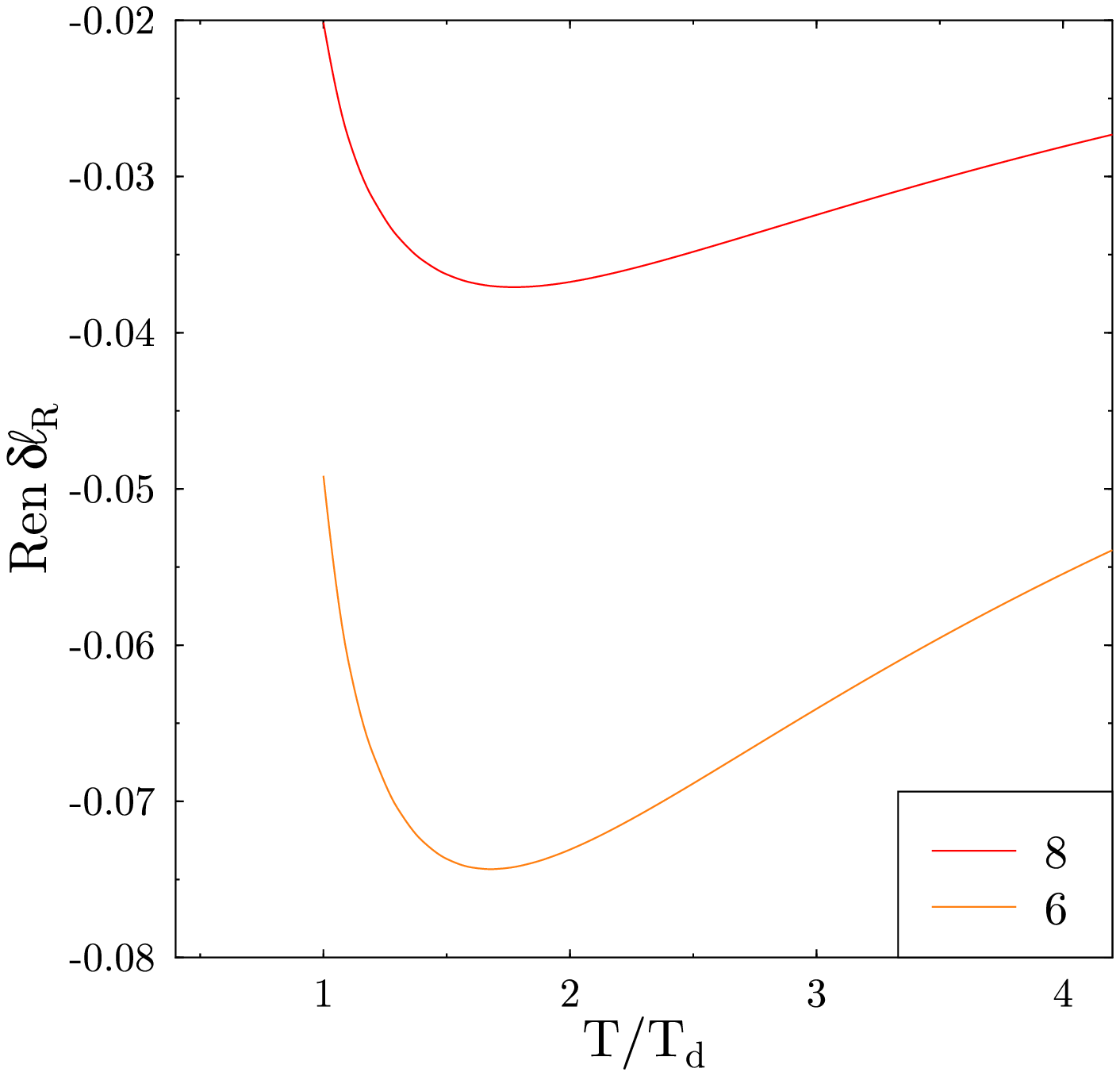}
\end{center}
\caption{Difference loops from the matrix model: Polyakov loops, minus 
their
large $N$ values.}
\label{fig:MMdiffloops}
\end{figure}

To obtain a more precise measure of the quality of our fits,
we computed the difference loops in our mean field approximation,
and plot them in Fig. \ref{fig:MMdiffloops}.  We do this
because even in this simple mean field approximation,
there are corrections to $N=\infty$ factorization at $N=3$.
Now compare the difference
loops in mean field theory, Fig. \ref{fig:MMdiffloops}, to those
from the lattice, Fig. \ref{fig:LoopsDiffs}.
As expected from general arguments, the adjoint difference
loop is always smaller than the sextet difference loop;
also, both difference loops are negative in mean field theory.
In detail, however, the difference loops found from mean field
theory are very different from those found from the lattice.
First, in magnitude the difference loops from mean field theory
are at least a factor of three times smaller than found from
the lattice.  Further, their temperature dependence is very
different: in mean field theory, both difference loops are
greatest at about $\approx 1.5 T_d$, with approximately the
same width, $\pm .5 T_d$.  
In contrast, the difference loops from the lattice have
a maximum much closer to $T_d$; while the sextet has a tail
which persists to $\approx 3 T_d$, the octet really appears to be a sharp,
narrow spike.

The quality of the fit could be improved by including
other terms in the action.  
We started by including an adjoint loop in the potential
at each site, (\ref{zN_potential}), $\sim \gamma_8 \ell_8$.
Within our numerical accuracy, this only appeared 
to produce a shift in $\beta_3 \rightarrow \beta_3 + \gamma_8$.
We show in the next section that this can be understood at infinite $N$.

To model the sextet and octet loops, it is necessary
to add corresponding fields at each site.  
In (\ref{eff_act_RRp})
there are two diagonal couplings, $\beta_{6,6^*}$
and $\beta_{8,8}$, and one off-diagonal coupling, $\beta_{3,6}$.
Even in mean field theory, it is tedious to solve numerically 
for several, coupled condensates; the results of a more
careful study will be presented separately \cite{coming1}.

The presence of other loops obviously feeds back into the
triplet loop.  One of the clearest tests of this is the
value of the triplet loop at $T_d$.  In a mean field
theory which includes just the triplet loop, (\ref{mean_field_3}),
numerically we estimate that 
$\ell_3(T_d) \approx .485 \pm .001$; this is consistent
with the value of $\approx .49$ in \cite{kogut}.
This is significantly higher than the values of the renormalized
loop from the lattice, where both we and \cite{bielefeld_ren}
find $\widetilde{\ell}_3(T_d) \approx .4$.  
In a $N=3$ matrix model which includes both triplet and
sextet loops, we find that the value of the triplet
loop at $T_d$ decreases significantly by adding
$\beta_{3,6}$, as this represents a linear coupling between the two
loops \cite{coming1}.
It is also possible to decrease $\ell_3(T_d)$
by adding a term for the decuplet loop to the potential $\cal W$ 
with the appropriate sign \cite{coming1}.

Thus the approximate linearity in $\beta_3$ with $T$,
(\ref{eq:beta3fit}), should be treated as preliminary.
Further, we doubt that it is true for all couplings.  
In particular, since the octet difference loop is such
a sharp, narrow spike in temperature, it appears that we
can only model it with a $\beta_8$ which varies non-monotonically
with temperature; {\it i.e.}, which is itself a spike.
This is not so obvious for the sextet loop, due
to the triplet-sextet mixing from the coupling
$\beta_{3,6}$.
The sextet loop is also affected by its coupling with itself, 
through the coupling $\beta_{6,6}$, and by the decuplet loop
in the potential.

Nevertheless, the matrix model proposed
in sec. IV.A appears to be a useful way of characterizing the
condensates of renormalized Polyakov loops when $N=3$ \cite{coming1}.

\subsection{Matrix Models: $N>3$}

For general $N$, the simplest possibility is to start with
an action including just the fundamental loop, 
\beq
{\cal S}_N = - \frac{N^2}{6} \beta \;
\Sigma_{i,\hat{n}} \; {\rm Re} \; \ellN(i) \ellN^*(i + \hat{n})
\; ,
\label{eff_act_N}
\eeq
$\beta \equiv \beta_{N,N^*}$.
For the action to be of order $N^2$ at large $N$, 
$\beta$ must be of order one as $N\rightarrow \infty$.
Positive values of $\beta$ correspond to a
ferromagnetic coupling.  As the perturbative
vacuum at high temperature is completely ordered, 
$\beta\rightarrow + \infty$ as $T\rightarrow \infty$.
We assume $\beta > 0$ \cite{zero_beta}.

At infinite $N$, if the fundamental loop condenses, 
loops in higher representations are fixed by factorization.
The condensate for the fundamental
loop is determined by the potential, $\cal W$ of (\ref{zN_potential}).

We start with the mean field analysis of (\ref{eff_act_N}),
following the discussion of Kogut, Snow, and Stone \cite{kogut}.
We then discuss how these results change when the potential
$\cal W$, (\ref{zN_potential}), is added \cite{mixed_action,mixed_action_2}.
A similar discussion was given recently by Aharony {\it et al} \cite{susy7}.

Replacing the values of all nearest neighbors by an average
value $\ell_0$, we need to evaluate the integral at one site,
$$
\calz = \int \; d\boldl_{ N}
\; \exp\left( N^2 (2 \beta \ell_0) \; {\rm Re} \; \ell_{ N} \right) 
$$
\beq
\equiv \exp\left(- N^2 {\cal V}_{GW}(\beta \ell_0)\right)\; .
\label{zss_N}
\eeq
The mean field condition is
\beq
\ell_0 = - \frac{1}{2 \beta}
\; \frac{\partial}{\partial \ell_0} {\cal V}_{GW}(\beta \ell_0) \; .
\label{mean_field_N}
\eeq
This condition is equivalent to minimizing the mean field
potential
\beq
{\cal V}_{mf}(\beta,\ell) = \beta \ell^2 + {\cal V}_{GW}(\beta \ell) \; .
\label{potential_mf}
\eeq
We replace $\ell_0$ by $\ell$, and interpret the
result as a potential for $\ell$.  
At a fixed $\beta$, as usual the vacuum $\ell_0$ is given
by minimizing ${\cal V}_{mf}$ with respect to $\ell$.

Since an overall factor of $N^2$ is scaled out of the potential,
a nonzero value of ${\cal V}_{mf}(\ell_0)$
implies that the free energy $\sim N^2$.  
This is expected in the deconfined phase from the liberation
of $\sim N^2$ gluons.  In the confined phase, 
Thorn \cite{susy2} noted that as all states are color singlets,
their free energy is at most of order one, so
${\cal V}_{mf}(\ell_0)\sim 1/N^2 \approx 0$.
This scaling also motivated the Polyakov loop model
\cite{pol_loop_a,pol_loop_b,pol_loop_c,cargese}.
Here, we assume that to go from the single site model to thermodynamics,
we multiply ${\cal V}_{mf}$ by $T^4$ times the volume of space.

The potential ${\cal V}_{GW}$
has been computed in the large $N$
limit by Gross and Witten \cite{gross_witten}.  
At infinite $N$ the result is
nonanalytic, and is given by two {\it different} potentials.
For small $\ell$, the potential is just a mass term,
\beq
{\cal V}_{mf}^- 
= \beta (1 - \beta) \ell^2 \;\;\; , \;\;\; \ell \leq \frac{1}{2 \beta} \; ,
\label{pot_mf_less}
\eeq
while at large $\ell$, the potential is 
\beq
{\cal V}_{mf}^+
= - 2 \beta \ell + \beta \ell^2 + 
\frac{1}{2} \log (2 \beta \ell) + \frac{3}{4} 
\;\;\; , \;\;\; \ell \geq \frac{1}{2 \beta} \; .
\label{pot_mf_greater}
\eeq

The physical interpretation of this potential is rather different
from the context in which it arose.
Gross and Witten considered a $U(N)$ lattice gauge
theory in two dimensions, with lattice coupling constant 
$\beta_{GW}\equiv \beta \ell$ \cite{gross_witten}.
This is the only parameter in the model, and there
is no condition to fix $\ell_0$.  Instead, 
the expectation value of ${\rm Re} \;\ell_{ N}$ is related to the string
tension; it changes with $\beta_{GW}$, but is always nonzero.
The two potentials in
(\ref{pot_mf_less}) and (\ref{pot_mf_greater}) correspond to
weak and strong coupling branches of the free energy.
About $\beta_{GW} =1/2$, the first and second
derivatives of the free energy
are continuous, but the third derivative is not, so there 
is a third order phase transition in $\beta_{GW}$.

In mean field theory, $\beta$ is an effective coupling for
the fundamental loop.
As a function of $\ell$ at fixed $\beta$,
the first and second derivatives of the potential are
everywhere continuous, but third (and higher) derivatives
are discontinuous at a single point, 
when $\ell = 1/(2 \beta)$.  This nonanalyticity
is special to $N=\infty$:
the mean field potential is everywhere continuous for finite $N$.

Overlooking this discontinuity, the potential
${\cal V}_{mf}$ behaves as a potential should.
When $\beta < 1$, the potential just increases monotonically
with $\ell$, so the minimum is at
$\ell_0 = 0$.  For $\beta > 1$, 
the potential about the origin is given by ${\cal V}^-_{mf}$,
and so starts out with a negative mass term.
The potential decreases 
with increasing $\ell$, with a single minimum when
\beq
\ell_0 = 
\frac{1}{2} \left( \; 1 \; + \; 
\sqrt{ \frac{\beta - 1}{\beta} } \; \; \right) \; .
\label{vev_ell}
\eeq
When $\ell > \ell_0$, 
the potential increases monotonically.
For all $\beta$, 
the potential is bounded at large $\ell$,
${\cal V}^+_{mf} \sim + \beta \ell^2$ as $\ell \rightarrow \infty$.
(For $\ell > 1/(2 \beta)$, the extremal condition 
$\partial {\cal V}^+_{mf}/\partial \ell = 0$ is a quadratic
equation.  There is another root besides (\ref{vev_ell}),
but it occurs for $\ell < 1/(2 \beta)$, and so doesn't matter.)

Thus there is a confined phase for $\beta < 1$, 
and a deconfined phase for $\beta > 1$.
The expectation value of the loop, $\ell_0$,
jumps from zero below the transition,
$\beta = 1^-$, to $\frac{1}{2}$ just above, $\beta = 1^+$.
The latter is (\ref{vev_Td}) of the Introduction.
In the limit $\beta \rightarrow \infty$, $\ell_0 \rightarrow 1$.

To verify that the transition is in fact of first order,
consider the value of the potential at its minimum.
In the confined phase, $\ell_0 = 0$, so ${\cal V}_{mf}^-(\ell_0) = 0$
for all $\beta < 1$, including $\beta \rightarrow 1^-$.  
In the deconfined phase, using (\ref{vev_ell}) one finds that
\beq
{\cal V}_{mf}^+(\ell_0) \approx 
- \; \frac{\beta - 1}{4} + \ldots \; ,
\label{order_mf}
\eeq
as $\beta \rightarrow  1^+$.  Thus 
the first derivative of ${\cal V}_{mf}(\ell_0)$,
which respect to $\beta$, is discontinuous when $\beta = 1$.  

If we assume that $\beta$ is linear in the temperature ---
as found for three colors --- then the deconfining
transition 
is thermodynamically of first order at $N = \infty$, with
a latent heat $\sim N^2$.  
In this we agree with \cite{kogut}.

Even so, the non-analyticity of the potential still has striking
physical consequences.
In particular, exactly at 
$\beta = 1$, the potential is completely {\it flat}
for $\ell$ between $0$ and $\frac{1}{2}$,
${\cal V}^-_{mf}=0$ \cite{flat}.  The
potential then increases monotonically,
starting out to cubic order
in $\ell - \frac{1}{2}$.  The only reason the
order parameter can jump, despite
the flatness of the potential, is because $\beta = 1$ is special:
then, and only then, does the point at which the potential is
discontinuous coincide with the nontrivial minimum.

To appreciate this in another way, consider 
how the mass squared, 
$m^2 = \partial^2 {\cal V}_{mf}/\partial \ell^2$
changes.
Approaching the transition in the confined phase
implies that we compute about $\ell_0=0$, 
\beq
m^2_- \approx 2 (1 - \beta) \;\;\; , \;\;\;
\beta \rightarrow 1^- \; .
\label{confined_mass}
\eeq
In contrast, approaching the transition in the deconfined phase,
we compute about $\ell_0 = \frac{1}{2}$, 
\beq
m^2_+ \approx 4 \sqrt{ \beta - 1} \;\;\; , \;\;\;
\beta \rightarrow 1^+ \; .
\label{deconfined_mass}
\eeq
Thus while both masses vanish at $\beta = 1$, they
vanish with different powers of $|\beta - 1|$.

The mass of the Polyakov loop is of physical significance
in the underlying gauge theory \cite{pol_loop_b,pol_loop_c}.
In coordinate space, the connected two point function of $\ell(\vec{x})$ is:
\beq
\langle \ell^*_{ N}(\vec{x}) \ell_{ N}(0)\rangle
- |\langle \ell_{ N}\rangle|^2 \sim
\frac{\exp(- M |\vec{x}|)}{|\vec{x}|} \;\;\; , \;\;\;
|\vec{x}| \rightarrow \infty \; .
\label{two_point_ell}
\eeq
In the confined phase, $M = \sigma/T$, where $\sigma$ is the string
tension.  In the deconfined phase, one can define $M = 2 m_{Debye}$,
where $m_{Debye}$ is a (gauge-invariant) Debye mass.
Assuming that $M \sim m$,
with mass dimensions made up by some other physical mass scale, 
such as the temperature, this mean field theory predicts that the string
tension vanishes at the transition as \cite{alvarez}:
\beq
\sigma(T) \sim (T_d - T)^{1/2}  \; \; \; , \;\;\;
T \rightarrow T_d^- \; ,
\label{string_tension_Td}
\eeq
and the Debye mass, as
\beq
m_{Debye}(T) \sim (T - T_d)^{1/4} \; \; \; , \;\;\;
T \rightarrow T_d^+ \; .
\label{debye_mass_Td}
\eeq
One might refer to this as a ``critical'' first order transition:
at the transition, the order parameter jumps, but
the masses vanish, asymmetrically.

We next include the effects of the potential, (\ref{zN_potential}).
We start with the simplest term, $\gamma_2 \neq 0$, which
is the contribution of the adjoint loop to the potential.
At large $N$, in mean field approximation we need to evaluate the integral
$$
\widetilde{\calz} = \int \; d\boldl_{ N}
\; \exp\left( N^2 \left( 2 \beta \ell \, {\rm Re}\, \ell_{ N} 
+ \gamma_2 |\ell_{ N}|^2 \right) \right)
$$
\beq
\equiv \exp\left(-N^2 \widetilde{{\cal V}}(\beta \ell,\gamma_2)\right) \; .
\label{adj_potential}
\eeq
The solution is \cite{mixed_action,mixed_action_2}:
\beq
\widetilde{{\cal V}}(\beta \ell,\gamma_2) = 
\gamma_2 k^2 + {\cal V}_{GW}(\beta \ell + \gamma_2 k) \; ;
\label{adj_potential_with_k}
\eeq
where $k$ is a variable which one minimizes with respect to.
The variation with respect to $k$ enforces the condition that the
expectation value of $\ell_{ N}^2$ satisfies factorization.
Finally, the mean field solution is given by minimizing a potential
with respect to $\ell$ and $k$:
\beq
\widetilde{{\cal V}}_{mf}(\ell,k) = 
\beta \ell^2 + \gamma_2 k^2 + {\cal V}_{GW}(\beta \ell + \gamma_2 k) \; .
\label{mixed_pot}
\eeq
This is trivial to solve.  Expanding about $k = \ell + \delta k$,
for $\delta k = 0$, this reduces to the previous mean field
potential, except that the coupling constant is shifted,
$\beta \rightarrow \beta + \gamma_2$.  
Further, if $\ell$ is
extremal with respect to this shifted mean field, the term
linear in $\delta k$ also vanishes.  This follows
because the Gross--Witten potential
is a function only of $\beta \ell$, and not of $\beta$ and $\ell$
separately.  As discussed previously, 
for $N=3$ we discovered numerically
that in mean field theory, we can shift
the adjoint coupling away, $\beta \rightarrow \beta + \gamma_2$.  

This is very different from lattice gauge theories in two dimensions,
where $\beta_{GW}$ and $\gamma_2$ represent independent coupling
constants \cite{mixed_action,mixed_action_2}.  Then $\beta_{GW}$
and $\gamma_2$ can be varied irrespectively of each other, and one
finds that the third order transition, in $\beta_{GW}$ for 
$\gamma_2 = 0$, can become a first order transition
in the plane of $\beta_{GW}$ and $\gamma_2$
\cite{mixed_action,mixed_action_2}.  

The generalization to $\gamma_4 \neq 0$ is direct.
One includes a constraint field for $|\ell|^2$ and then solves
the constraint at $N=\infty$.  The mean field
coupling $\beta$ is again shifted, 
$\beta \rightarrow \beta + \gamma_2 + \# \gamma_4 \ell_0^2 $,
but by an amount which depends on the condensate, $\ell_0$;
there are also additional potential terms, $\sim \gamma_4 \ell_0^4$.  
Consequently, $\gamma_4 \neq 0$ corresponds to a change in the
potential.  Aharony {\it et al.} \cite{susy7} show that for
$\gamma_4 \neq 0$, ordinary transitions appear to
be generic.  In particular, first order transitions have finite
correlation lengths in both phases.

This remains the case for an arbitrary potential, $\cal W$.
At large $N$, this is just a sum of powers of the fundamental loop:
\beq
{\cal W} = \Sigma_{i,m} \; 
\left(
\gamma_{2m} \;
(|\ell_N|^2)^m \; + \;
\xi_{m N} \left( \left( \ell_N \right)^{m N} + {\rm c.c.} \right)\right) 
\; .
\label{potential_infinite_N}
\eeq
This is a typical potential for a scalar field $\ell_N \equiv \ell_N(i)$:
the adjoint loop acts like a mass term, 
while the other terms are interactions of quartic and higher order,
invariant under a global symmetry of $Z(N)$.

We define the Gross--Witten point as the transition for
a potential which is just the adjoint loop.  All other
interactions are dropped:
$\beta + \gamma_2 \neq 0$, with
$\gamma_4 = \gamma_6 = \ldots = \xi_N = \xi_{2N} = \ldots = 0$.

The deconfining transition for three colors appears to be close to
the Gross-Witten point of infinite $N$.  
For example, although the data are very limited \cite{string_ten},
the decrease of the Debye mass near $T_d$ does seem to be
significantly sharper than for the string tension, as indicated
by the different ``critical exponent'' in (\ref{debye_mass_Td})
versus (\ref{string_tension_Td}).
Exactly how close $N=3$ is to the Gross--Witten point of $N=\infty$ can
be characterized within matrix models \cite{coming1}.  
Effects which are important for three colors include the
contribution of the decuplet loop, which is like a cubic interaction
for the triplet loop, 
and the mixing between the triplet and sextet loops, $\beta_{3,6}$
in (\ref{eff_act_RRp}).

Assuming that three colors is near the Gross--Witten point, we can
explain why $Z(N)$ neutral fields have small expectation values
in the confined phase, (\ref{small_neutral}).  In the confined
phase, the potential is purely a mass term, 
as corrections to (\ref{pot_mf_less}) 
begin with the baryon vertex, $\sim (\ell_N)^N$ 
\cite{kogut,largeN_trick,matrix_review}.
This vertex induces $Z(N)$ neutral expectation values,
but as noted by Goldschmidt \cite{largeN_trick},
these are of order $\sim \exp(- N)$.

That $N=3$ is close to the Gross--Witten point 
could be an accident of three colors.
The lattice will tell us if 
the deconfining transition for four or more colors is also close
to the Gross--Witten point.
The lattice finds a first order transition \cite{four_colors,teper},
but the crucial tests are
the value of the renormalized, fundamental loop at $T_d^+$,
and whether the string tension and the Debye mass
decrease significantly near $T_d$.

If this is not found, the most probable scenario is just that
the transition becomes more strongly first order with increasing
$N$.  We term the transition strongly first order if the value of 
the renormalized fundamental loop at $T_d^+$ is near unity.
Then at the transition, deconfinement is not halfway, as 
it is at the Gross--Witten point, but nearly complete.

For such a strongly first order transition, neither
the string tension, nor the Debye
mass, would need to change much about $T_d$.
Gocksch and Neri \cite{gocksch_neri} argued that
the ``loop'' string tension is constant for $T \leq T_d$.  
In a Nambu string model, though, a large $d$ expansion
shows that when the ordinary
string tension vanishes at $T_d$, the loop string tension
is still $\approx 87\%$ of its value at zero temperature \cite{alvarez}.
We are unaware of lattice data on the loop string tension.

\section{Outlook}

In this paper we presented a general analysis of the renormalization
of Polyakov loops, and applied it to measure the simplest loops
for three colors in four dimensions.  These results led us to
consider effective matrix models for the deconfining
phase transition.  
There are clearly many avenues for future study.

For two colors, there will be two regions.  About $T_d$, there
is a critical region, controlled by universality of the second
order transition \cite{sv_yaffe}, and in which factorization fails.
This may then match onto a mean field region, where factorization
is approximately valid.  

For three colors, 
careful measurements of the renormalized loops,
and associated masses, 
will sharply constrain the couplings of the effective matrix
model \cite{coming1}.
We note that while we could not extract an expectation
value for the renormalized decuplet
loop from that for the bare loop, mean field theory indicates that
it is significant.

Simulations should quickly show if for the
deconfining transition for four or more colors
is near the Gross--Witten point.  

Considering theories other than $SU(N)$,
Holland, Minkowski, Pepe, and Wiese \cite{G2}
noted that in a pure $G(2)$ gauge theory,
there is no center to the gauge group, and so no absolute
measure of confinement.
This is analogous, though, to $Z(N)$ neutral loops in $SU(N)$, 
for which we measured no signal below $T_d$.
In the simplest mean field theory for a $G(2)$ gauge theory, 
presumably 
there is a first order transition, with a value for the fundamental
loop at $T_d^+$ near $\frac{1}{2}$. 
Maybe like $SU(3)$, the ``deconfining'' transition
in a $G(2)$ gauge theory is also near the Gross--Witten point.

The renormalization of Wilson lines implies that 
once the divergent mass is known,
{\it all} renormalized loops can be computed.  We suggest
that numerical simulations measure loops of different
shapes, such as Polyakov loops with cusps, Fig. (\ref{fig:cusp}),
and circular loops, as can be computed in supersymmetric
theories \cite{susy6}.

In the end, however, what is most important is to
measure renormalized Polyakov loops for theories with dynamical
quarks.  Our method for computing renormalized Polyakov loops
is completely unaffected by the presence of dynamical quarks.
Given the flavor
independence found for the pressure \cite{flavor_ind}, 
it would be striking if 
the values of renormalized Polyakov
loops, with dynamical quarks, are found to be close to those
of the pure gauge theory.  For recent results, see \cite{bielefeld_ren}.

{\bf Acknowledgements:}
The research of A.D. is supported by
the BMBF and the GSI; Y.H. and
R.D.P., by U.S. Department of Energy grant DE-AC02-98CH10886;
J.T.L., by D.O.E. grant DE-FG02-97ER41027;
K.O., by the RIKEN/BNL Research
Center and D.O.E. grant DF-FC02-94ER40818.
The numerical work was done on a cluster of
workstations at Brookhaven National 
Laboratory using the publicly available MILC code.
J.L. thanks P. Arnold and H. Thacker for useful discussions.
R.D.P. thanks: R. Brower and C. P.
Korthals-Altes, for emphasizing the importance of mixed actions
in matrix models; M. Creutz, for numerous discussions
on group theory and the lattice; P. Damgaard, for many
discussions of his work; G. Korchemsky, for explaining
loops with cusps; also, Y. Dokshitzer,
N. Drukker, D. J. Gross, T. Heinzl, and S. Necco.
R.D.P. and Y.H. thank P. Petreczky for discussions concerning
Polyakov loops on the lattice.

\appendix*
\section{Improved Wilson Line}

It is difficult extracting renormalized Polyakov loops in representations
such as the decuplet, because the bare loop is suppressed by a small
renormalization constant.  
In this Appendix we give a formal discussion of how
to improve the Wilson line \cite{improved}.

Our discussion applies to any Wilson line along
a path $x^{\mu}(s)$, where $s$ is the path length along the curve.
As shown in sec. II, (\ref{propagator}),
the propagator for a test quark is proportional 
to the Wilson line.  Consequently, we consider a generalized
propagator, by adding an operator ${\cal X}$
to the covariant derivative:
\beq
\left( \frac{d}{ds} - i g A^\mu \dot{x}^\mu - {\cal X} \right)
\; {\cal G}(s,s') \; = \; \delta(s-s') \; ;
\label{definition_G}
\eeq
where $\dot{x}^\mu = dx^\mu/ds$.  The representation is
denoted implicitly.  Schematically, the solution to this equation is
\beq
{\cal G}(s,s') = \theta(s-s')\;
{\cal P}\; \exp \left( \int ( i g A^\mu \dot{x}^\mu + 
{\cal X} ) ds \right) \; ,
\label{solution_G}
\eeq
with $\cal P$ denoting path ordering.  The solution to ${\cal G}$
is schematic because of the path ordering, but it is easy to understand
the solution as a power series in ${\cal X}$, with each
insertion of ${\cal X}$ sandwiched between a Wilson line on both
sides.

Any possible operator ${\cal X}$ has a higher mass dimension
than the gauge field, so we make up the dimensions with inverse powers
of the ultraviolet cutoff, such as the lattice spacing $a$.
The important thing is that $\cal X$ respects the relevant
symmetries.  Gauge invariance requires that
$\cal X$ transforms homogeneously under gauge
transformations, but that is simply done by using powers of
the field strength tensor, $G_{\mu \nu}$.  
The operator must also be invariant under how we parameterize
arc length, $s \rightarrow s'(s)$.  The final symmetry is
the zig-zag symmetry of Polyakov \cite{zig_zag}; for
reparameterizations which go in the opposite direction,
with $ds/ds' < 0$, ${\cal X}$ should change sign.  

The simplest possibility is
\beq 
{\cal X} = a^2 g^2 \; G^2_{\mu \nu} \sqrt{\dot{x}^2} \; .
\eeq
This is gauge covariant and reparameterization invariant,
but is not zig-zag symmetric.  If one abandons zig-zag symmetry,
then this operator can be used to regularize the Wilson line,
since it is just a field strength dependent ``mass'' term for the line.  

One can satisfy all symmetries with the following operator,
\beq 
{\cal X} = a g \; 
G_{\mu \nu} \frac{\dot{x}^\mu \ddot{x}^\nu}{\dot{x}^2} \; ,
\eeq
where $\ddot{x} = d^2 x^\mu/ds^2$.  A similar
operator was noted by Polyakov and Rytchkov \cite{zig_zag}.
Since it vanishes for a straight path, where $\ddot{x}^\mu = 0$,
it doesn't help to regularize the Wilson line.  
It is also problematic to continue to Minkowski spacetime, since
it is singular on the light cone \cite{zig_zag}.  

To define an operator which satisfies all of our requirements, we
define a unit vector normal to the path, 
\beq
\hat{n}_\mu \dot{x}^\mu = 0 \;\;\; , \;\;\;
\hat{n}^2 = 1 \; .
\eeq
For a given direction, we introduce the $\hat{n}$ dependent operator
\beq
{\cal X}_{\hat{n}} = \kappa g \; G_{\mu \nu} \dot{x}^\mu \hat{n}^\nu \; .
\eeq
We then define a modified Wilson line as
\beq
\int d\Omega_{\hat{n}}
\; {\cal P}\; \exp \left( \int ( i g A_\mu \dot{x}^\mu + 
{\cal X}_{\hat{n}} ) ds \right) \; .
\eeq
The operator ${\cal X}_{\hat{n}}$ inserts the field strength
tensor perpendicular to the path.  We then integrate over
all directions of the insertion, with 
$d\Omega_{\hat{n}}$ the normalized integral over $\hat{n}$,
$\int d\Omega_{\hat{n}} = 1$.
We obviously cannot integrate over all directions $\hat{n}$
in the exponential, or the term would vanish, and so
do so in the prefactor of the exponential.
Zig-zag symmetry is maintained by integrating
over all $\hat{n}$.

In perturbation theory, to lowest order ${\cal X}_{\hat{n}}$ generates
new divergences.  In three spatial dimensions, and including a Debye
mass $m_{D}\sim gT$ in the propagator for $A_0$ \cite{gava}, 
the leading divergence is
\beq
\sim \frac{g^2}{T} 
\int d\Omega_{\hat{n}}
\int \frac{d^3 k}{(2 \pi )^3}
\; \left( \frac{- 1 + a^2 \kappa^2 (\hat{n}\cdot \vec{k})^2}{k^2 + m_D^2} 
\right) \; .
\label{ren_loop_kappa}
\eeq
We assume that the ultraviolet divergence is cutoff strictly at momenta $1/a$.
The usual term is $\sim \int d^3k/k^2 \sim 1/a$,
and is of the same order as the new term,
$\sim a^2 \int d^3 k \sim 1/a$.
For $\kappa < 1/\sqrt{3}$, 
(\ref{ren_loop_kappa}) is negative, while for 
$\kappa > 1/\sqrt{3}$, it is positive.  If a more realistic cutoff is
used, then the value of $\kappa$ at which the sign changes will be
different, but for large $\kappa$ (and to leading
order in $g^2$) bare loops are enhanced, not suppressed.

Following the procedure in sec. (II.D), from
$\log(\langle \ell \rangle)$ we compute as power series
in $N_t$.  The term linear in $N_t$ gives $\calz_\calr$,
with the renormalized loop given by the term independent of $N_t$.
We argue that while the $\calz_\calr$ are $\kappa$ dependent,
the renormalized loops are not, at least in perturbation theory.
To lowest order in $g^2$, for $\kappa = 0$
the renormalized loop arises from the correction from the
Debye mass term, 
$\sim g^2 (m_D^2)^{1/2}/T \sim g^3$, (\ref{pert_loop}) \cite{gava}.
This is non-analytic in the Debye mass, and arises because the
leading term is only linearly divergent.  For the $\kappa$-dependent
term, the leading term is cubically divergent, so corrections
are $\sim g^2 \kappa^2 a^2 m_D^2/(a T) \sim g^4 \kappa^2 aT$; but
this is $\sim 1/N_t$, and vanishes as $N_t\rightarrow \infty$.
There is a non-analytic term at one higher order in momentum,
but this is even smaller, 
$\sim g^2 \kappa^2 a^2 (m_D^2)^{3/2}/T \sim g^5 \kappa^2 (aT)^2
\sim 1/N_t^2$.
It seems likely that this holds for any $\kappa$-dependent terms:
they contribute to terms $\sim 1/(a T) \sim N_t$, or to
lattice corrections $\sim 1/N_t$ or smaller, but not to terms
in the continuum limit, $\sim N_t^0$.  This analysis is special to four
spacetime dimensions: in three dimensions, the $\kappa$ dependent
term is quadratically divergent, and has logarithmic corrections,
as for $\kappa = 0$.

While zig-zag symmetric, the term added, ${\cal X}_{\hat{n}}$,
is not a phase factor; the coupling $\kappa$ must be real in
order to enhance the bare loop.  As the bare loop is
not the trace of a unitary matrix, there is no bound on the
renormalized loop, as for $\kappa = 0$, (\ref{constraint});
when $\kappa \neq 0$, $\calz_\calr$ can diverge in
the continuum limit, instead of vanishing.  

On the lattice, adding a field strength tensor at a point
corresponds to stapling a plaquette, as the average
over $\hat{n}$ becomes a sum over the directions transverse to
the path.  
To lowest order in $\kappa$, this modification is the same as
the smearing of link variables proposed by
the APE collaboration \cite{improved}.
We suggest doing this not just to lowest order, but
to all orders in $\kappa$.  This is related to the
stout links of Morningstar and Peardon \cite{improved}.
The difficulty is that eventually the Wilson line is smeared
over the entire lattice, which must then be cut off in some way.
On the other hand, the usual problem with smeared links 
is the need to project back to an element of $SU(N)$,
which is unnecessary here.

\newpage


\begin{thebibliography}{999}
\bibitem{zero_beta}
If $\beta_N = 0$, couplings
for the two-index tensor, and adjoint, representations, must be included,
$\beta_{N(N \pm 1)/2}$ and $\beta_{N^2 - 1}$, respectively.
(For $N\geq 4$, both symmetric and anti-symmetric two-index
tensor representations enter,
with corrections to factorization $\sim 1/N$.)
The case where the couplings for
$Z(N)$ charged fields vanish,
 $\beta_N = \beta_{N(N \pm 1)/2} = 0$, but
that for the $Z(N)$ neutral field(s) do not, 
such as $\beta_{N^2 - 1} \neq 0$, is special.  In this case,
the global $Z(N)$ symmetry is promoted to a local symmetry,
and while $Z(N)$ neutral fields can condense above some
temperature, loops with nonzero $Z(N)$ charge cannot at $N< \infty$.
We assume this doesn't happen: that 
$\beta_N \neq 0$, and other couplings are small corrections.
\bibitem{flat}
The flatness of ${\cal V}_{mf}$ at $\beta = 1$ was
noted by \cite{kogut} and \cite{green_karsch}, but
dismissed as unphysical.  At the time, however, neither
the decrease in the string tension near $T_d$,
nor the weakness of the first order transition, were seen
in numerical simulations \cite{string_ten}.
\end{thebibliography}
\end{document}